\documentclass{aa}  

\usepackage{amsmath, bm}
\usepackage{tikz,pgfplots}
\usepackage{multirow}
\usepackage{multirow}
\usepackage{graphicx,array}
\usepackage{caption}
\usepackage{subcaption}
\usepackage[linkcolor=blue]{hyperref}

\graphicspath{{images/}}

\def\bpsi{{\bm \Psi}}
\def\btheta{{\pmb \theta}}
\def\bphi{{\pmb{\phi}}}

\def\loss{{\mathcal L}}
\newcommand{\argmin}{\operatornamewithlimits{argmin }}

\def\a{{\bf a}}
\def\bb{{\bf b}}
\def\e{{\bf e}}
\def\m{{\bf m}}
\def\x{{\bf x}}
\def\y{{\bf y}}

\def\R{\textit{R}}

\def\P{{\bf P}}

\def\W{{\bf W}}

\def\pyse{PySE}
\def\aegean{AEGEAN}
\def\pybdsf{PyBDSF}
\def\reim{Re \& Im}
\def\magphas{Mag \& Phase}
\def\sota{state-of-the-art}
\newcommand{\acosmos}{A$^{3}$COSMOS}

\usepackage{txfonts}

\begin{document}

   \title{Challenging interferometric imaging: Machine learning-based source localization from uv-plane observations}


   \author{O. Taran
          \inst{1}
          \and
          O. Bait
          \inst{2}
          \and
          M. Dessauges-Zavadsky
          \inst{2}
          \and    
          T. Holotyak
          \inst{1}
          \and
          D. Schaerer
          \inst{2}          
          \and
          S. Voloshynovskiy
          \inst{1}
          \fnmsep\thanks{S. Voloshynovskiy is a corresponding author}
          }

   \institute{Department of Computer Science, University of Geneva, Switzerland \\
            \email{olga.taran@unige.ch, svolos@unige.ch}
            \and
            Observatoire de Gen\`eve, Universit\'e de Gen\`eve, 51 Chemin Pegasi, 1290 Versoix, Switzerland\\
            \email{daniel.schaerer@unige.ch}
            }


 
  \abstract
   {Rising interest in radio astronomy and upcoming projects in the field is expected to produce petabytes of data per day, questioning the applicability of traditional radio astronomy data analysis approaches under the new large-scale conditions. This requires new, intelligent, fast, and efficient methods that potentially involve less  input from the domain expert.}
   {In our work, we examine, for the first time, the possibility of fast and efficient source localization directly from the uv-observations, omitting the recovering of the dirty or clean images.}
   {We propose a deep neural network-based framework that takes as its input a low-dimensional vector of sampled uv-data and outputs source positions on the sky. We investigated a representation of the complex-valued input uv-data via the real and imaginary and the magnitude and phase components. We provided a comparison of the efficiency of the proposed framework with the traditional source localization pipeline based on the state-of-the-art Python Blob Detection and Source Finder (PyBDSF) method. The investigation was performed on a data set of 9164 sky models simulated using the Common Astronomy Software Applications (CASA) tool for the Atacama Large Millimeter Array (ALMA) Cycle 5.3 antenna configuration.}
   {We investigated two scenarios: (i) noise-free as an ideal case and (ii) sky simulations including noise representative of typical extra-galactic millimeter observations. In the noise-free case, the proposed  localization framework demonstrates the same high performance as the state-of-the-art PyBDSF method. For noisy data, however, our new method demonstrates significantly better performance, achieving a completeness level that is three times higher for sources with uniform signal-to-noise (S/N) ratios between 1 and 10, and a high increase in completeness in the low S/N regime. Furthermore, the execution time of the proposed framework is significantly reduced (by factors $\sim 30$) as compared to traditional methods that include image reconstructions from the uv-plane and subsequent source detections.}
   {The proposed framework for obtaining fast and efficient source localization directly from uv-plane observations shows very encouraging results, which could open new horizons for interferometric imaging with existing and future facilities.}


   \maketitle
%
\section{Introduction}

Radio astronomy is at the cusp of a revolution in terms of the sensitivity that can be achieved at centimeter and meter wavelengths. The various radio-astronomical interferometers  such as the LOw-Frequency ARray \citep[LOFAR;][]{LOFAR}, MeerKAT radio telescope \citep[][]{MeerKAT}, Australian square kilometer array pathfinder \citep[ASKAP;][]{ASKAP}, and Murchison Widefield Array \citep[MWA;][]{MWA} are already producing promising results. At the same time, radio astronomy is an extremely data-intensive science. It is expected that the upcoming projects will produce data volumes on an exabyte scale \citep{Scaife20}. Thus, it will be very challenging for astronomers to undertake standard radio data analysis tasks such as calibration, imaging and source localization using traditional techniques. Hence, it is vital to design fast and efficient techniques to replace the traditional data analysis approaches in radio astronomy.

At millimeter (mm) wavelengths the Atacama Large Millimeter Array (ALMA)  has led to several large imaging and spectroscopic programs thanks to its excellent sensitivity. Of particular interest in the context of the current work are the various large programs targeting extra-galactic deep fields. This includes, for example, the Reionization Era Bright Emission Line Survey \citep[REBELS;][]{REBELS}, the ALMA SPECtroscopic Survey in the Hubble Ultra-Deep Field \citep[ASPECS; ][]{ASPECS}, the ALMA Large Program to INvestigate [CII] at Early times \citep[ALPINE][]{ALPINE1, ALPINE2, ALPINE3}, and the GOODS-ALMA survey at 1.1 mm \citep{GOODS-ALMA}. Another rich ALMA data set on which our current study is focused is the Automated Mining of the ALMA Archive in the COSMOS Field (\acosmos) data set\footnote{\url{https://sites.google.com/view/a3cosmos/home?authuser=0}} \citep{liu19a3cosmos}. These surveys study the gas and dust properties of galaxies at high redshifts.

A key technical component while analyzing these data is to accurately identify sources (their positions, fluxes, and sizes) in noise-limited images produced by radio interferometers. This is important, for example, when calculating the number density and luminosity function of astronomical sources. It has important implications for constraining various physical models in astrophysics and cosmology. Traditionally, radio source properties were measured using a two-dimensional (2D) Gaussian fit to the light profile \citep{Condon97} and usually required some level of manual intervention. Such manual interventions become increasingly difficult when dealing with large-area ALMA surveys mentioned above. Also with the all-sky radio surveys such as the Faint Images of the Radio Sky at Twenty-cm (FIRST) survey \citep{FIRST-survey} and NRAO Very Large Array (VLA) Sky Survey \citep[NVSS;][]{NVSS}, which contain millions of sources, manual source detection is difficult. Future large-area radio surveys \citep{Norris13} are expected to detect   increase in the number of sources by an order of magnitude (e.g., the  Evolutionary Map of the Universe (EMU) survey \citep{Norris11}). Radio surveys of extra-galactic deep fields using upgraded radio facilities, such as the VLA-COSMOS 3 GHz survey \citep{Smolcic17} and the MeerKAT International GHz Tiered Extra-galactic Exploration (MIGHTEE) survey \citep{MIGHTEE_survey, Heywood22}, are already detecting several thousands of sources in a single field. 

Thus, it is essential to build an automatic source detection algorithm that would accurately detect sources above the noise level with high completeness and simultaneously have a low number of false detections. There are several automatic source detection tools available in the literature, for example, Search and Destroy (SAD)\footnote{\url{http://www.aips.nrao.edu/cgi-bin/ZXHLP2.PL?SAD}}, Source-Extractor
\citep[SExtractor;][]{Bertin96}, AEGEAN \citep{Hancock12, Hancock18} and  Python Blob Detection and Source Finder \citep[PyBDSF;][]{mohan2015pybdsf}, PySE \citep{carbone2018pyse}, CAESAR \citep{riggi2019caesar, riggi2016automated}, PROFOUND \citep{robotham2018profound, hale2019radio}, and SOFIA \citep{serra2015sofia, westmeier2021sofia}. \citet{Hopkins15} provide a detailed comparison between different source detection algorithms available in the literature and their limitations. This works very well for bright sources. However, it is much more interesting to detect faint sources, typically with a signal-to-noise ratio (S/N) below 5.0, with high level of completeness and purity. This not only offers the ability to probe sources at even higher redshifts than currently achieved. It also has the potential to detect new kinds of astronomical sources lying close to the noise level, which may have been missed by studies using traditional source detection techniques.

Recent advances in machine learning and in particular deep neural networks in the form of convolutional neural networks (CNNs) have led to a lot of success in radio astronomy. In particular, CNNs have been extensively used to classify radio galaxy morphologies
\citep[e.g.,][]{Aniyan17, Lukic19a, Ma19, Tang19, Bowles21, Riggi22}. In particular, \citet{Riggi22} have used Mask R-CNN object detection framework on the ASKAP EMU survey data to perform both object detections and classifications. \citet{Schmidt22} used CNNs designed for super-resolution applications directly on the UV data to up-sample features in the case of sparse sampling, for instance, very-long-baseline interferometry. The images produced from these UV data show good recovery of source properties. Also, CNNs have shown great promise in point source detection, as demonstrated in ConvoSource \citep{Lukic19b} and DeepSource \citep{DEEPSOURCE}. DeepSource is shown to be perfect in terms of purity and completeness down to a S/N of 4 and  outperforms the current state-of-the-art source detection algorithm PyBDSF in several metrics. Recently, it was shown that an encoder-decoder based neural network DECORAS \citep{DECORAS} can perform source detection even on dirty images down to a S/N of 5.5 and can also recover various source properties such as fluxes and sizes quite accurately. \citet{delli_veneri_22} further shows that deep learning-based source detection can also be performed on dirty spectral data cubes and shows a good recovery of source properties such as morphology, flux density, and projection angle.

A common problem in most of the source-finding algorithms in the literature is that they are performed on the image plane and mostly on CLEAN images, which are computationally expensive to produce and scale poorly with the data volumes. Thus, even if these source finding algorithms are made efficient, they will still be limited by the time taken to produce the CLEAN images. It is well known that CLEAN leads to several imaging artifacts that can affect the purity of these source-finding algorithms. And despite automating several steps in source finding, these approaches still require some amount of manual intervention to exclude imaging artifacts. In this work, we circumvent these problems by designing a novel direct uv-plane based source localization algorithm using recent advances in deep neural networks.

Motivated by the achievements of the deep neural networks (DNN) made in many domains, we propose a fast and efficient DNN-based framework for source localization. In general, this framework takes as input a low-dimension vector of sampled uv-observations and outputs the source position on sky in the form of a binary map. Our proposed framework is targeted towards source detections in ALMA continuum images at mm wavelengths, particularly with the aim of detecting low S/N sources and to speed up the  source detection process overall. We trained and tested our proposed framework on simulated ALMA data. We investigated the impact of different factors on the performance of the proposed framework, namely, the representation of complex-valued input data via real and imaginary or magnitude and phase real-valued components, the impact of receiver noise and atmospheric noise due to the presence of water vapor, the impact of the S/N values of the sources, and the impact of the number of sources in the field of view. We  provide a detailed analysis of the execution time of the proposed framework and compare it with those of the traditional source localization pipeline.

\begin{table*}[ht]
    \centering
    \renewcommand{\arraystretch}{1.5}%
    \caption{Summary of the data simulation parameters using CASA.}
    \begin{tabular}{ccp{0.35\linewidth}} \hline
        Parameter & Value & Notes \\ \hline 
        Antenna configuration & ALMA Cycle 5.3 & \\
        Number of antennas & 50 & \\
        Field & COSMOS & J2000 10h00m28.6s +02d12m21.0s \\                                                        Number of pointings or simulations  & 9164 & Every pointing is randomly chosen within a radius of 1 deg around the field. \\ \hline
        Central frequency  & Band 6 (230 GHz) &  \\
        Number of channels & 240 & \\
        Channel width      & 7.8 MHz & \\
        Sampling time & 10 secs &  \\
        Total Integration time & 20 mins & \\
        Hour angle & Transit & \\ \hline 
        Sky model dimensions & $512 \times 512$ pixel with 240 channels & \\
        Pixel size & 0.1\arcsec{} & \\
        Source type & Gaussian & The size of the major/minor axis and position angle is varied randomly and is chosen to be below the synthesized beam.\\
        Major (minor) axis & 0.4\arcsec{} to 0.8\arcsec{} & Typical resolution is 0.89\arcsec{} $\times$ 0.82\arcsec{} \\
        Position angle (PA) & 0 to 360 deg &  Chosen randomly  \\
        Number of sources & $0 - 5$ & Randomly chosen between 1 and 5. We add a few source free simulations  \\
        Source positions & random & Sources are randomly distributed within the primary beam. \\
        Flux range & 0.05 mJy to 0.5 mJy & The flux of each source is randomly chosen from this range assuming a uniform distribution. This roughly keeps a flat S/N range for our data set. \\
        Spectral index & 0 & We set this parameter to zero since the fractional bandwidth is quite small ($0.8\%$). \\ \hline 
        Primary beam & 22.86 \arcsec{} & \\
        Synthesized beam & 0.89\arcsec{} $\times$ 0.82\arcsec{} & \\
        Noise (\textsc{pwv}) & 1.796 & This parameter adds the receiver and atmospheric noise due to water vapor to the visibilities.\\
        Weighting & Natural & \\
        Robust & 0.5 & \\
        RMS noise (in images) & $\sim$ 50 $\mu$Jy & \\ \hline
    \end{tabular}
    \label{tab:simulation parameters}
\end{table*}

The paper is organized as follows. The ALMA data simulation procedure, the chosen parameters and the analysis are given in Sect. \ref{sec:data}. The traditional pipeline of source localization is described in Sect. \ref{sec:traditional pipeline}. The framework proposed in this work is explained in Sect. \ref{sec: proposed framework}. Section \ref{sec:results and analysis} is dedicated to the analysis of the obtained results and Sect. \ref{sec:discussion} offers comparisons with the literature. Finally, we present our conclusions in Sect. \ref{sec:conclusions}.

\section{Data}
\label{sec:data}

We train the proposed DNN-based framework on synthetic data with known source localization. All reported experiments have been performed exclusively on synthetic data. In the future, we plan to apply the proposed frameworks to the real observations taken from the (\acosmos) data set \citep{liu19a3cosmos}. Thus, our simulated  ALMA observations are designed to somewhat match the \acosmos~ observations.

\begin{figure}[t!]
    \centering
    \begin{subfigure}[c]{.32\linewidth}
      \centering
      \includegraphics[width=1.15\linewidth]{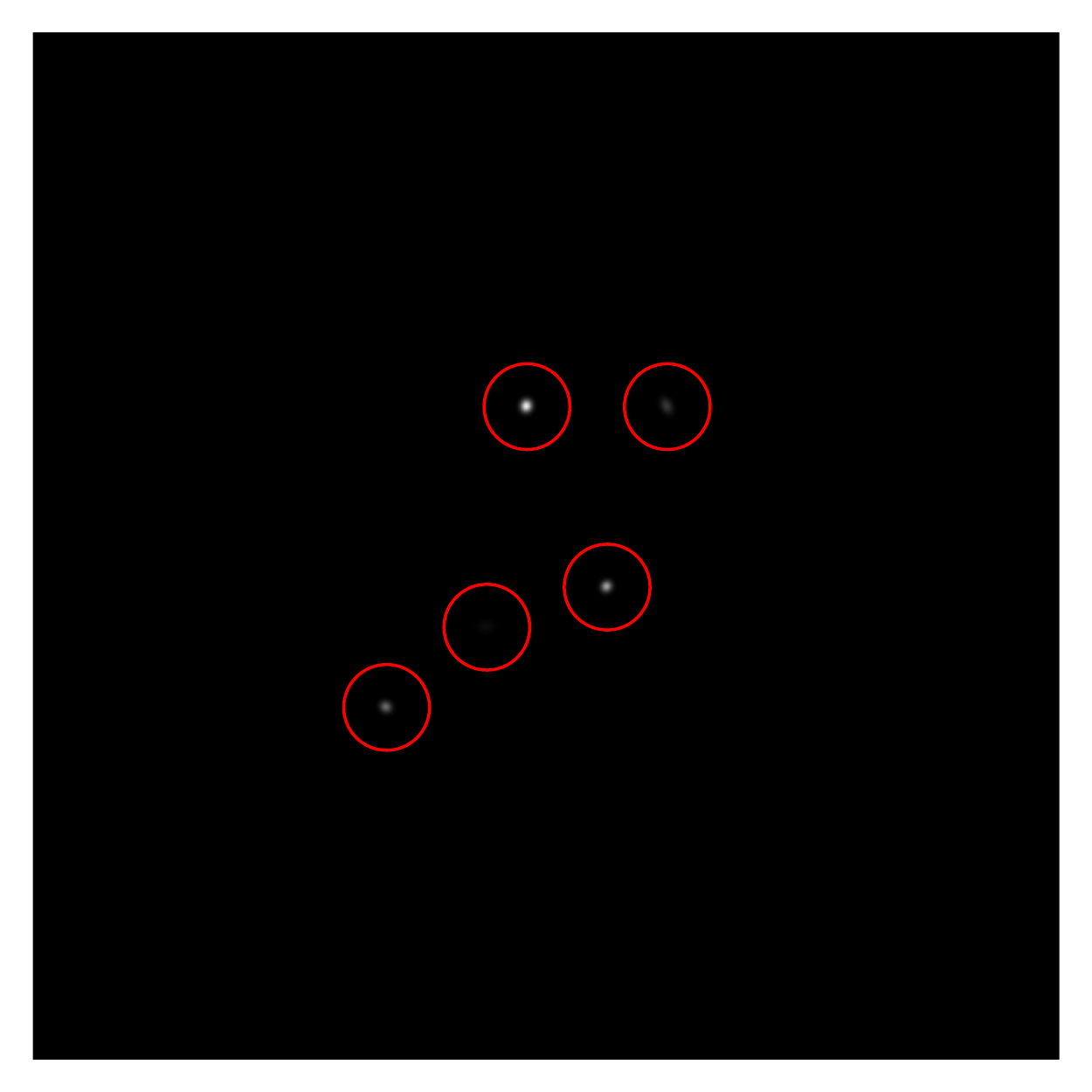}
    \end{subfigure}
    \hfill
    \begin{subfigure}[c]{.32\linewidth}
      \centering
      \includegraphics[width=1.15\linewidth]{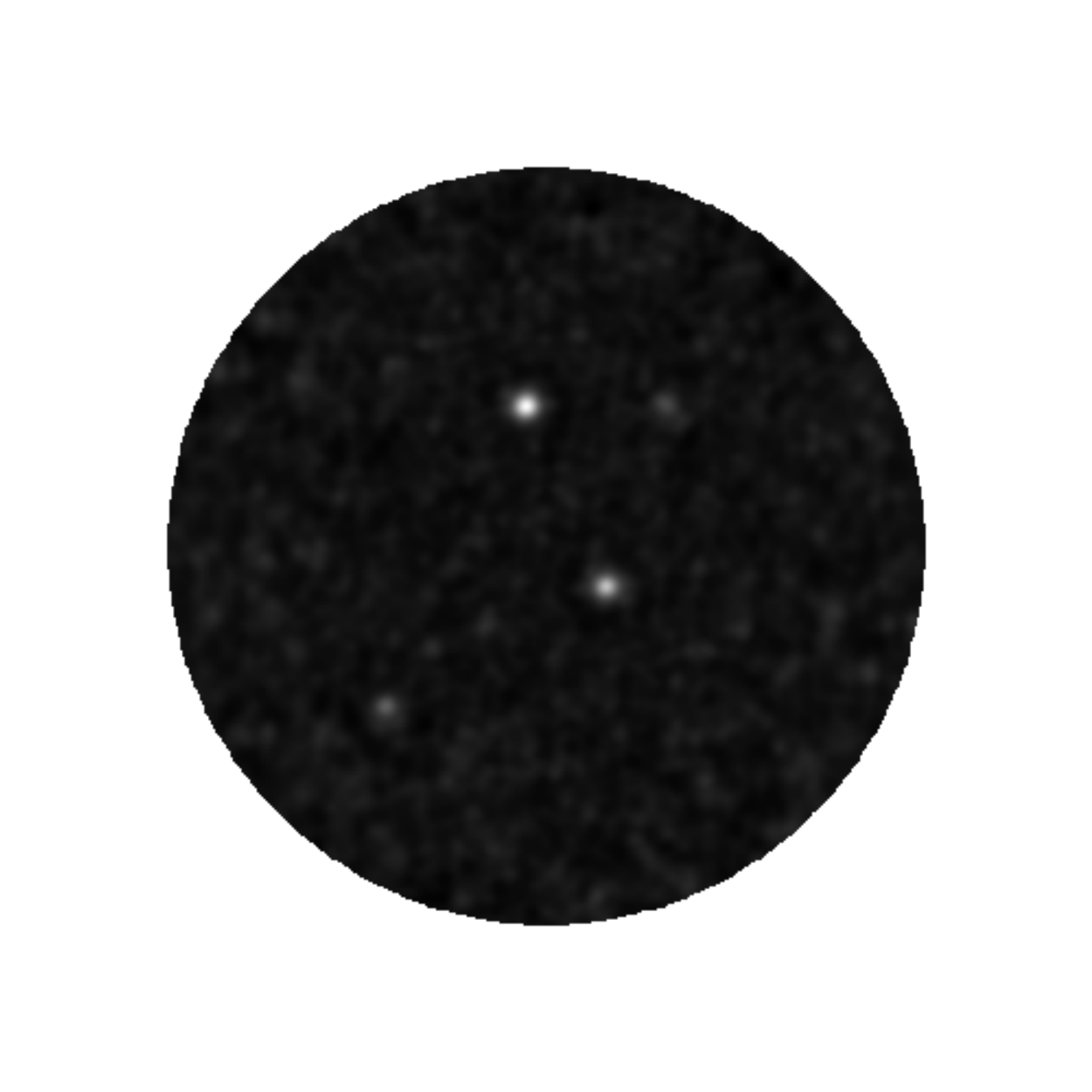}
    \end{subfigure}    
    \hfill
    \begin{subfigure}[c]{.32\linewidth}
      \centering
      \includegraphics[width=1.15\linewidth]{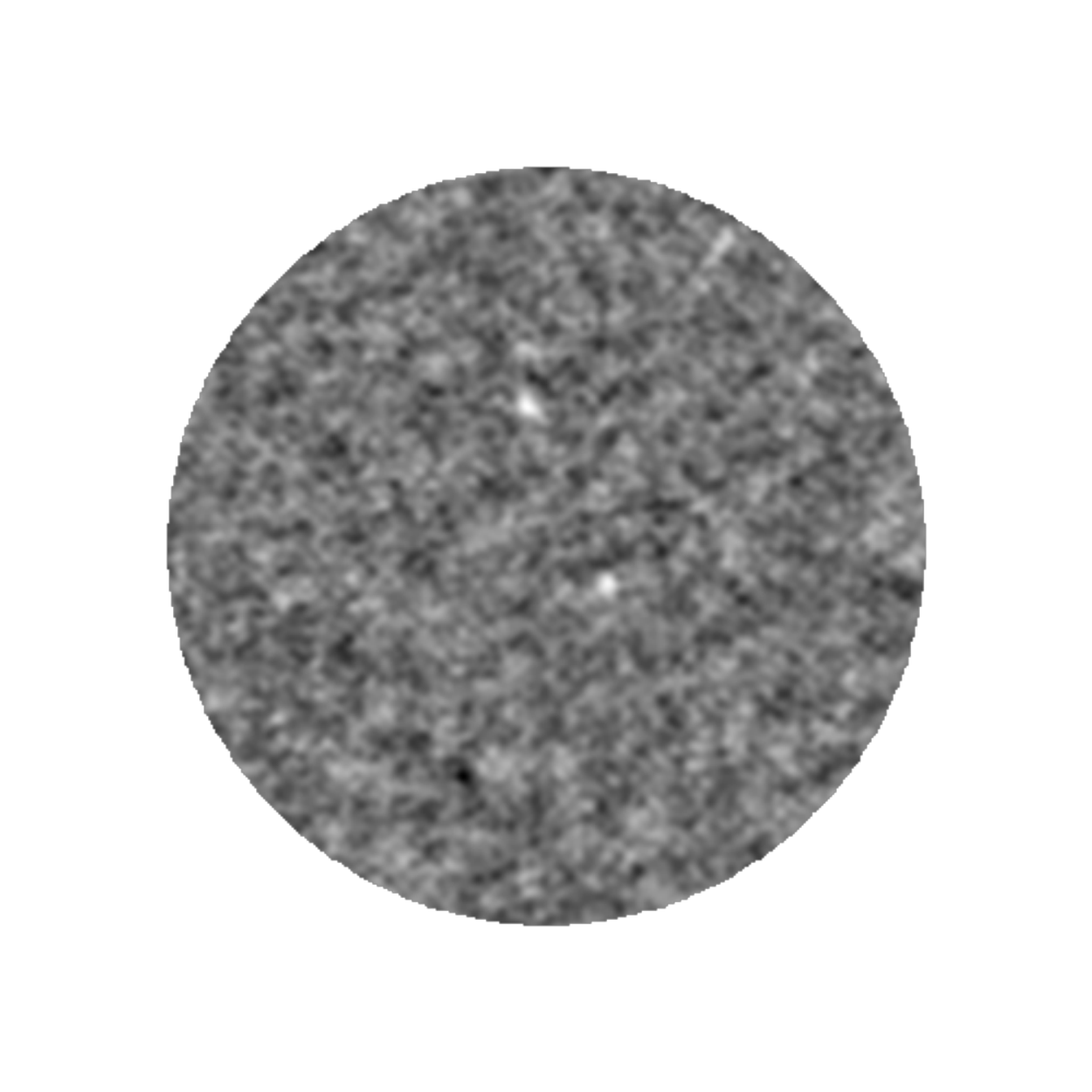}
    \end{subfigure}  
    \\
    \begin{subfigure}[c]{.32\linewidth}
      \centering
      \includegraphics[width=1.15\linewidth]{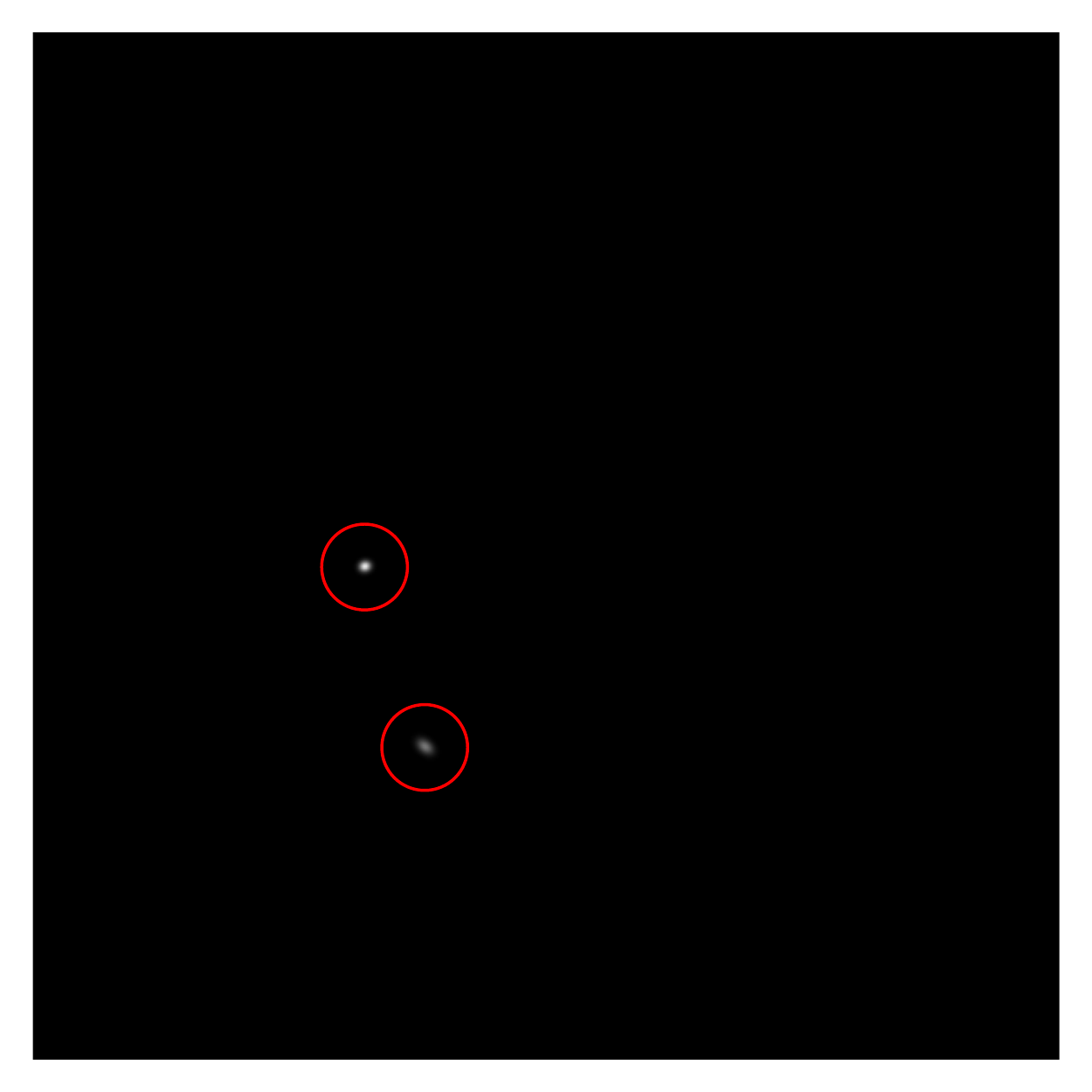}
    \end{subfigure}
    \hfill
    \begin{subfigure}[c]{.32\linewidth}
      \centering
      \includegraphics[width=1.15\linewidth]{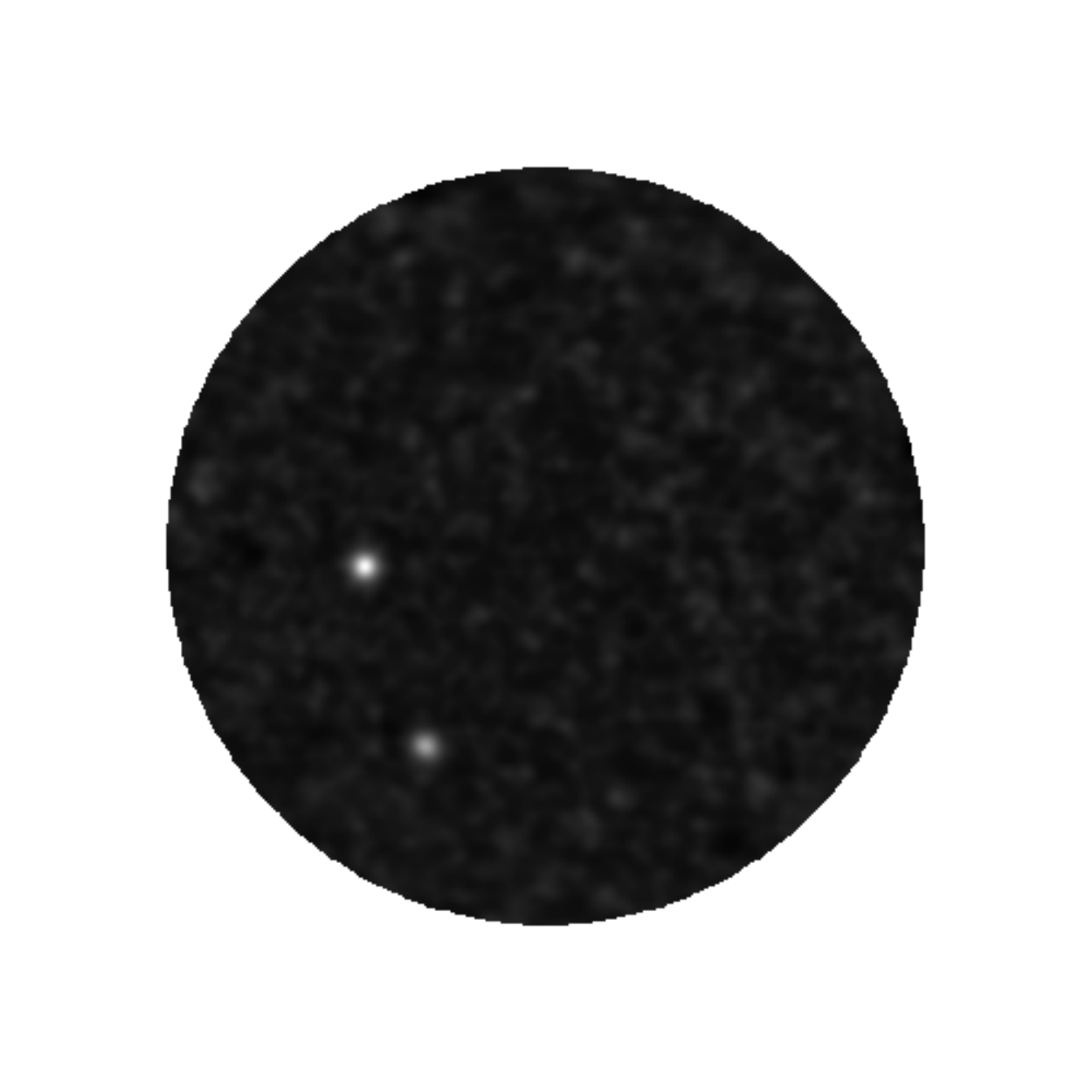}
    \end{subfigure}    
    \hfill
    \begin{subfigure}[c]{.32\linewidth}
      \centering
      \includegraphics[width=1.15\linewidth]{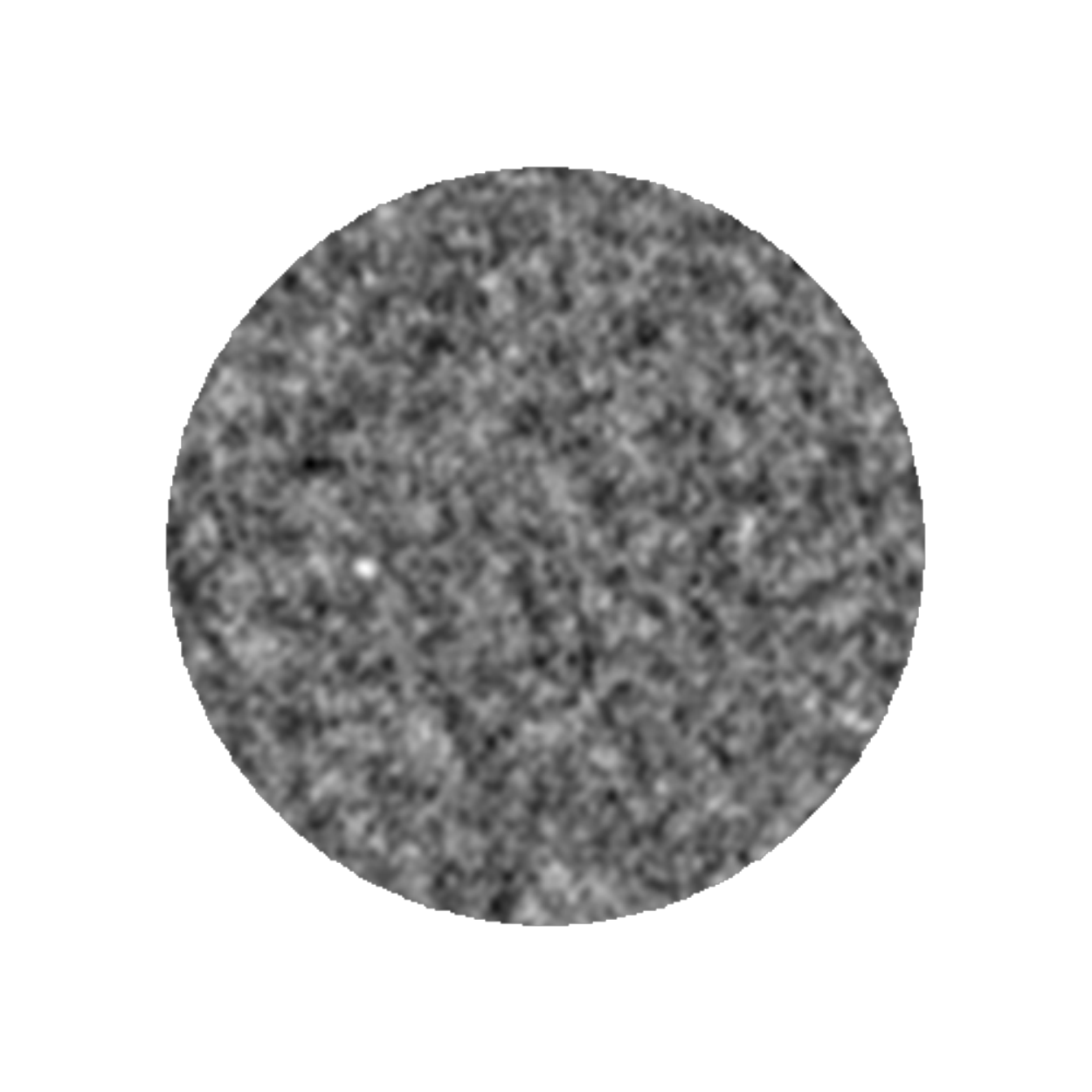}
    \end{subfigure}  

    \caption{Examples of simulated data used in this study. Top-left:\ True sky model. Top-middle:\ Noise-free model. Top-right:\ Noisy. Bottom-left:\ True sky model. Bottom-middle:\ Noise-free. Bottom right:\ Noisy. For both noise-free and noisy cases, the CLEAN representations are visualized. The sources are highlighted as red circles.}
    \label{fig:simulation examples}
\end{figure}

\begin{figure}[t!]
    \centering
    \begin{subfigure}[c]{.49\linewidth}
      \centering
      \includegraphics[width=0.65\linewidth]{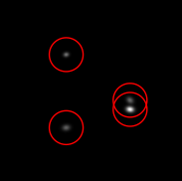}
    \end{subfigure}
    \hfill
    \begin{subfigure}[c]{.49\linewidth}
      \centering
      \includegraphics[width=0.65\linewidth]{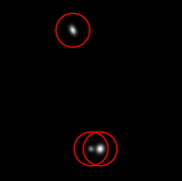}
    \end{subfigure}    
    \caption{Example of closely located sources of different intensity. For better visibility, the cropped and zoomed parts of the true sky models are shown.}
    \label{fig:close sources}
\end{figure}

For all our simulations, we used the Common Astronomy Software Applications (CASA) data processing software \textsc{v6.2} \citep{mcmullin2007casa, CASA}. We produced simulations for the 12-m ALMA array with a fixed configuration from the ALMA cycle 5.3. All our simulation pointings were randomly distributed within a radius of 1 deg around the COSMOS field centered on: J2000 10h00m28.6s +02d12m21.0s. For simplicity, we fixed the ALMA observing band to Band-6 centered at 230 GHz split in 240 channels with a channel width of 7.8 MHz. Our simulation pipeline follows the standard approach of first simulating a true sky model at a known phasecentre with known source positions and fluxes. The sky model consists only of sources with a Gaussian light profile with sizes close to or below the synthesized beam. In each pointing, the source position, size (i.e., the major and minor axis of the Gaussian), position angle, S/N, and total number of sources  (in the range between 1 and 5) are kept at random and drawn from an uniform distribution. This range in the number of sources per pointing and their sizes are chosen from true data from the \acosmos{} data in Band-6 for the ALMA compact configuration (similar to ours). Instead of drawing our simulation parameters exactly from the observed distribution in \acosmos{} data, we used a uniform distribution but the range of these parameters is motivated by the \acosmos{} data set (see the discussion in the following paragraph).
The sky model contains a $512 \times 512$ pixel image with a pixel size of 0.1\arcsec. This  sky model is used to generate the noise free uv-data using the \textsc{simalma} task \citep{mcmullin2007casa}. We sample each visibility every 10 secs for a total of 20 mins of integration on-source and choose the hour angle range such that the observing field is at transit. Appendix \ref{appendix: uvcoverage} shows a typical UV sampling for one of the simulations and the corresponding dirty beam. We then added the ALMA receiver and atmospheric noise due to water vapor to the uv-data using the \textsc{pwv} parameter value of 1.796 \citep{mcmullin2007casa}. This is a typical value in Band-6 as mentioned in the ALMA technical handbook\footnote{\url{https://almascience.nrao.edu/proposing/technical-handbook}}. We then average the visibilities,  both noisy and noise-free, in time and frequency, by gridding the visibilities on a uniform grid using the \textsc{msuvbin} task \citep{mcmullin2007casa}. Finally, we produce the dirty and CLEAN images from both the noisy and noise-free visibilities using the standard \textsc{tclean} task \citep{mcmullin2007casa}. For our simulation setup, we typically reach a rms noise of 50 $\mu$Jy in our dirty and CLEAN images. The size of the primary and synthesized beam is 22.86\arcsec and 0.82\arcsec, respectively. Table \ref{tab:simulation parameters} summarizes the various simulation parameters. 

Although geared towards  real observations from \acosmos, our simulations are intentionally chosen to not match them exactly{}. In particular, our sources are assigned random on-sky positions, fluxes, and sizes -- as opposed to choosing directly from the distribution of sources from true observations or those motivated by cosmological simulations. The phase center and number of sources in each simulation pointing are also randomly distributed around the COSMOS field. This ensures that during training we do not end up in the region of latent space of our deep-learning model, which is only trained to work  for a particular data set. Thus, it does not learn additional patterns which might exist in real observations or cosmological simulations such as, source clustering, the luminosity function, size distribution etc. Instead, our model is trained to be flexible enough to be applied to any other new ALMA data. On the other hand, for our current work, we fix the ALMA observing band, channel width and total integration time. This is to reduce the number of free parameters in ALMA simulations. In the future, we will test the effect of changing these two parameters especially when we apply our model to real data.

\begin{table}[t!]
   \renewcommand{\arraystretch}{1.5}%
    \centering
    \caption{Number of sky models with different numbers of sources in the test subsets.}        
    \begin{tabular}{ccccccc} \hline
        & \multicolumn{5}{c}{Number of sources} & \multirow{2}{*}{Total} \\
                   & 1 & 2 & 3 & 4 & 5 &  \\ \hline
    Test subset 1  & 161 & 177 & 196 & 177 & 205 & 916 \\
    Test subset 2  & 197 & 191 & 163 & 182 & 183 & 916 \\
    Test subset 3  & 165 & 175 & 181 & 180 & 215 & 916 \\ \hline    
    Average        & 174 & 181 & 180 & 180 & 201 & 916 \\
    \end{tabular}

    \label{tab:test sets statistics}
\end{table}

We produced a total of 9164 independent simulations or pointings with the sky model containing between 1 to 5 sources. In total, we ended up with 27632 sources across all the pointings. Each simulation produces $\sim$ 3.5 GB of data which includes the uv-data and dirty and CLEAN images that is $\sim$ 35 TB for the entire data set\footnote{The produced data set is available by request.}. The simulations are produced on the LESTA-computing cluster. The more details are given in the Appendix \ref{apppendix: lesta}.

Figure \ref{fig:simulation examples} and \ref{fig:close sources} show examples of the simulated data. The red circles highlight the positions of the sources in the true sky models. It is important to mention that the generated data set includes the same challenges encountered in real data, for example, the presence of blended sources (as in Fig. \ref{fig:close sources}) and sources that are hardly distinguishable from the background noise as in Fig. \ref{fig:simulation examples} (top-middle and bottom-right).
For the training of our framework, we perform the train-validation-test splitting three times with different seeds. Table \ref{tab:test sets statistics} shows the number of sky models with different numbers of sources in the test subsets. 

\begin{figure*}[t!]
    \centering
    \includegraphics[width=1\linewidth]{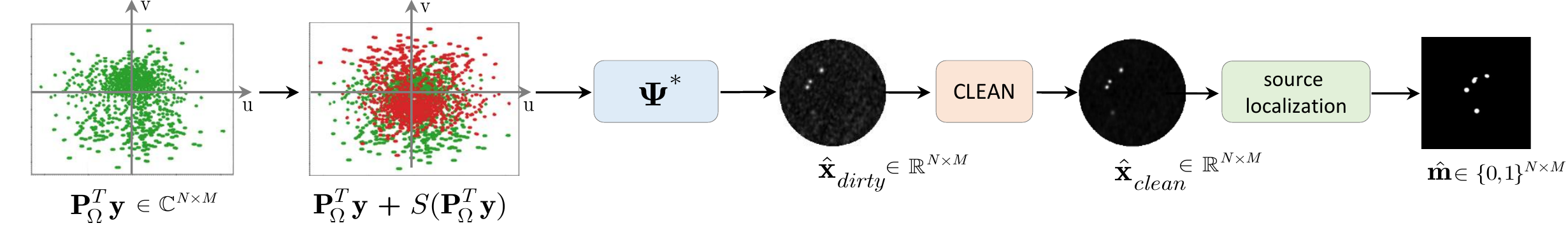}
    \caption{Schematic representation of the traditional pipeline.} 
    \label{fig:classical_pipeline}
\end{figure*}

\vspace{0.75cm}

The notation we use is defined as follows: $\x \in \mathbb{R}^{N \times M}$ denotes the true sky model, where $N$ and $M$ are the image size, set to 512 in our simulations. Then, $\hat{\x}_{dirty} \in \mathbb{R}^{N \times M}$ and $\hat{\x}_{clean} \in \mathbb{R}^{N \times M}$ denote dirty and CLEAN image, respectively. $\y \in \mathbb{C}^K$ stands for the sampled uv-data with $K = 1400$\footnote{The technical details are given in Appendix \ref{apppendix: dnn sampling}.}. $\m \in \{0, 1\}^{N \times M}$ is a binary source map, in which source pixels are set to 1, and background pixels are set to 0.
We calculated the S/N as:
\begin{equation}
     \textrm{S/N} = \frac{\textrm{total flux}}{\sigma_{\textrm{noise}}},
    \label{eq:snr}
\end{equation}
where the total flux is the true source intensity and $\sigma_{\textrm{noise}}$ is the noise rms. As explained in Table \ref{tab:simulation parameters}, $\sigma_{\textrm{noise}} \sim 50 \mu$Jy.  


\section{Traditional pipeline of source localization}
\label{sec:traditional pipeline}

The traditional pipeline of source localization is based on the reconstruction of the dirty and CLEAN images. The schematic representation of the traditional pipeline is given in Fig. \ref{fig:classical_pipeline}.

\subsection{Dirty image recovering}

The recovering of the dirty image might be formulated as a recovering of a high dimension sky model\footnote{The sky model $\x$ is of size $N \times M = 512 \times 512$. For the simplicity of notations we use the vectorized representation $N \cdot M$.}  $\x \in \mathbb{R}^{N \cdot M}$ from the corresponding low-dimensional sampled uv-visibility  $\y \in \mathbb{C}^{K}$ {\small($K << N \cdot M$)} corrupted by noise $\e$:
\begin{equation}
    \y = \W\x + \e, 
\end{equation}
%
where $\W =\P_\Omega \bpsi \in \mathbb{C}^{K \times N\cdot M}$ is a measurement sub-sampling matrices with an orthonormal basis of $\bpsi \in \mathbb{C}^{N\cdot M \times N \cdot M}$ and sampling operator of $\P_\Omega : \mathbb{C}^{N\cdot M} \to \mathbb{C}^{K}$ and $|\Omega| = K$.

It should be pointed out that due to the physical imaging constraints in the radio-astronomy, $\bpsi$ corresponds to the Fourier operator and $\P_\Omega$ is determined by the antennas configuration, measured frequencies, Earth's movement, sampling, and integration time. 
    
The recovering of the $\hat{\x}_{dirty}\in \mathbb{R}^{N \times M}$ consists in: (i) $\P^T_\Omega\y$ - expanding the observation $\y$ to a $N \times M$ representation by placing zeros in the entries corresponding to $\Omega^C$ that is the complementary support set of $\Omega$, (ii) adding the corresponding  symmetrical signal via $S(.)$ related to the symmetry property of Fourier and then (iii) applying the inverse Fourier $\bpsi^*$: 
\begin{equation}
    \centering
    \hat{\x}_{dirty} = \bpsi^*\Big( \P^T_\Omega\y + S\big(\P^T_\Omega\y \big) \Big).
\end{equation}

\begin{figure*}[ht!]
    \centering
    \includegraphics[width=1\linewidth]{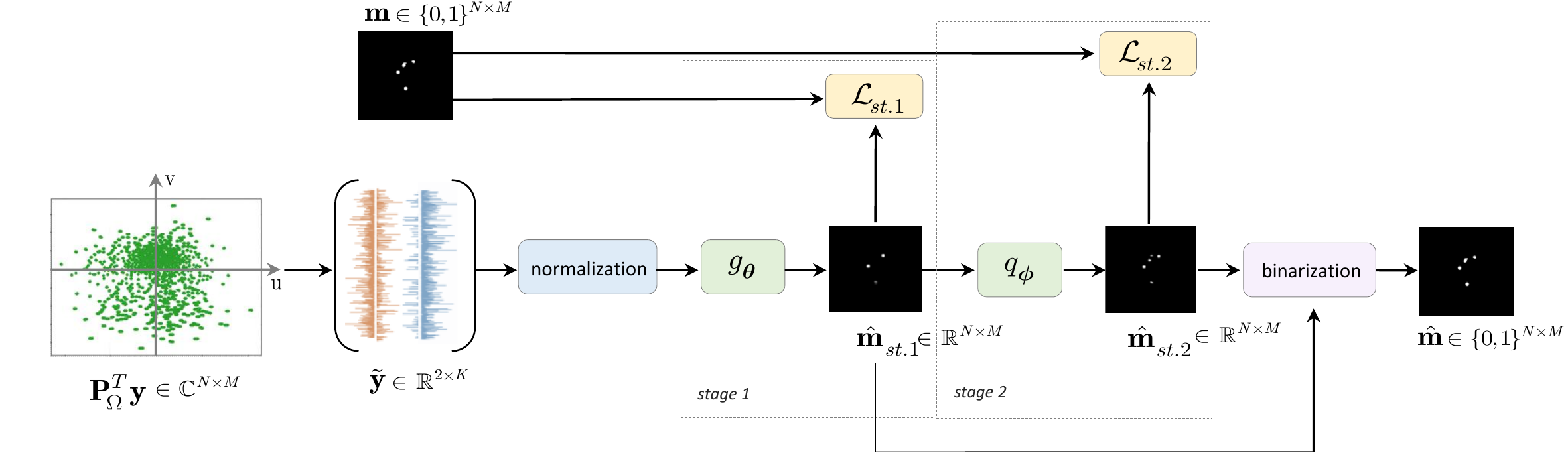}
    \caption{Schematic representation of the proposed framework.}
    \label{fig:main model}
\end{figure*}

\begin{figure*}[t!]
    \centering
    \begin{subfigure}[c]{.49\linewidth}
      \centering
      \includegraphics[width=1\linewidth]{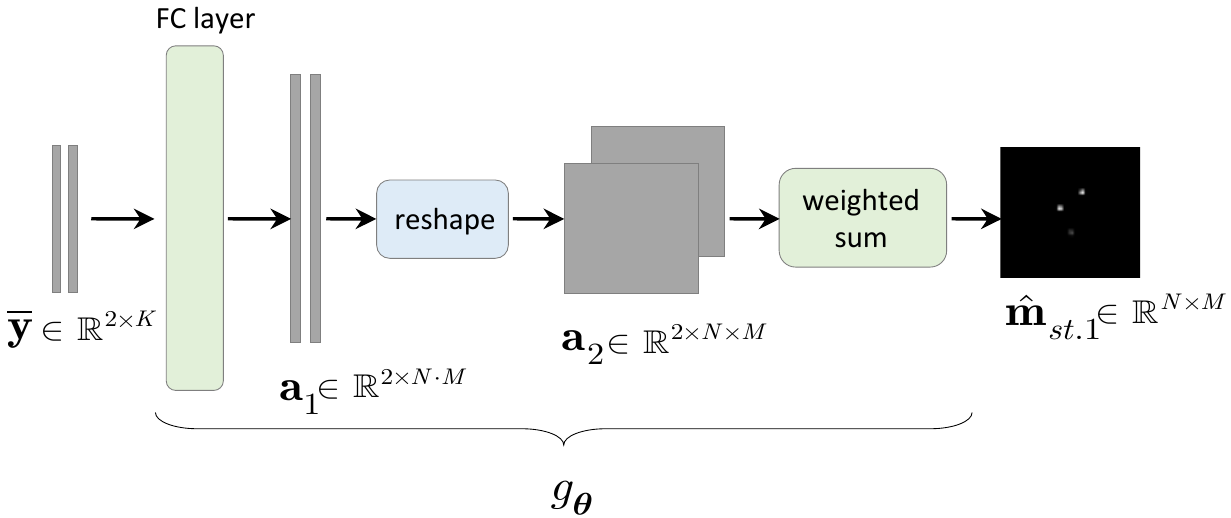}
    \end{subfigure}
    \hfill
    \begin{subfigure}[c]{.49\linewidth}
      \centering
      \includegraphics[width=1\linewidth]{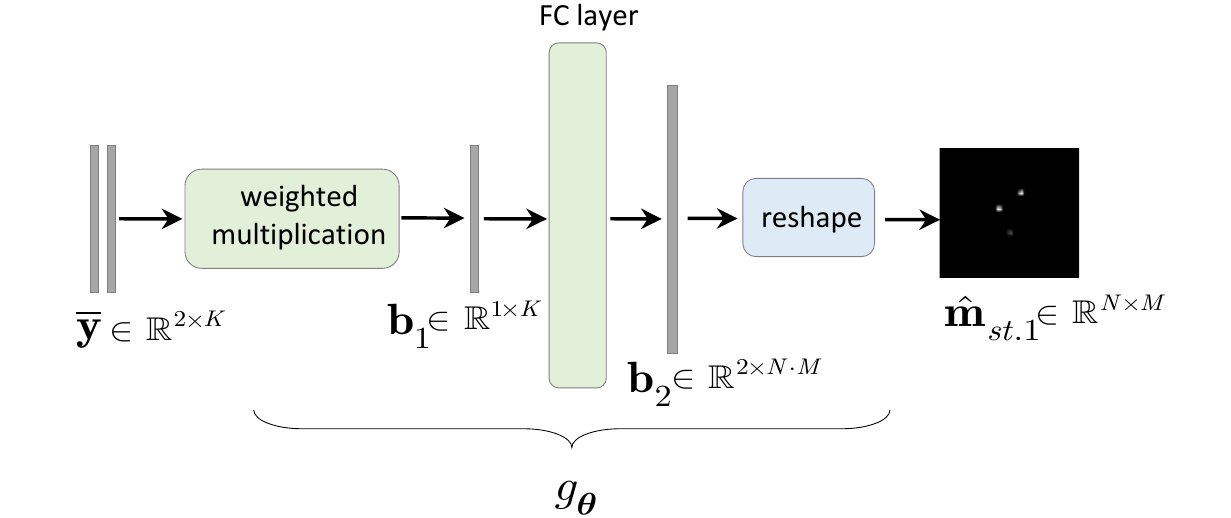}
    \end{subfigure}    
    \caption{Detailed architecture of the $g_\btheta$ model, where $\bar{\y}$ denotes the normalized input data $\Tilde{\y}$: in the left panel, $\bar{\y} \in \mathbb{R}^{2 \times K}$ is represented by the real and imaginary components. In the right panel, $\bar{\y} \in \mathbb{R}^{2 \times K}$ is represented by the magnitude and phase components.}
    \label{fig:stage 1 g}
\end{figure*}

\subsection{Clean image recovering}

Due to the presence of missing Fourier frequencies in $\P_{\Omega^C}$, the dirty image contains a lot of artifacts, leading to an increase in the false detected sources. In order to remove these artifacts, the CLEAN algorithm \citep{hogbom1974aperture} is applied to the dirty image. 

The CLEAN fundamental method is used in radio astronomy and consists of several steps. First, it finds the intensity and position of the peak that is of the greatest absolute intensity in the dirty image $\hat{\x}_{dirty}$. Second, it generates at this position a spike of an intensity equal to the product of a damping factor and the intensity at that position. Usually, the damping factor is $\le$ 1 and is termed the loop gain. The generated spikes are convolved with the instrumental point source function (PSF). Then, the obtained instrumental response is subtracted from the dirty image $\hat{\x}_{dirty}$. This procedure is repeated unless any remaining peak is below some user-specified level. The search for peaks may be constrained to specified areas of the image, called CLEAN windows. Then the accumulated point sources are convolved with an idealized CLEAN beam that, usually, is an elliptical Gaussian fitted to the central lobe of the dirty beam. Finally, the remaining residual of the dirty image is added. The obtained reconstruction is called CLEAN image $\hat{\x}_{clean}$.

In terms of the fast processing of large amounts of data, CLEAN as well as its accelerated versions such as the $w$-Stacking Clean \citep[WSCLEAN;][]{offringa2014wsclean}  act as the main bottleneck. The reconstruction remains time-consuming since several parameters have to be adjusted in iterative runs.

\subsection{Source localization}

Traditionally, the source localization is applied to the CLEAN images. Depending on the antenna configuration and the corresponding amount of artifacts, it might  also be applied to the dirty images. Nowadays,  a broad range of different source localization approaches exist that are used in the traditional pipeline. \cite{Hopkins15} provide a good overview of these methods. In our work, we focus on PyBDSF \citep{mohan2015pybdsf}. 


\section{Proposed framework}
\label{sec: proposed framework}

We propose a DNN-based framework that performs the source localization in the form of a binary map $\m \in \{0, 1\}^{N \times M}$ directly from the uv-data by taking  only sampled visibility data without reconstruction of dirty or CLEAN images as an input. The general scheme of the basic framework is illustrated in Fig. \ref{fig:main model} and consists of three steps: (1) input data pre-processing (i.e., normalization), (2) DNN-processing: stage 1 and 2, (3) post-binarization  and source localization. 

It should be pointed out that the sampled visibility data $\y \in \mathbb{C}^K$ are complex values. However, modern DNNs are not designed to work with complex values. To deal with this, we decompose the complex-valued uv-samples $\y \in \mathbb{C}^{K}$ into real-valued real and imaginary or magnitude and phase representation $\Tilde{\y} \in \mathbb{R}^{2 \times K}$. Due to the nature of the Fourier transform, in the case of real and imaginary representation, the trained model $g_\btheta$ at  the stage 1 is additive in nature, as shown in on the left side of Fig. \ref{fig:stage 1 g}. While in case of the magnitude and phase, it is multiplicative in nature as shown on the right side of Fig. \ref{fig:stage 1 g}. The other steps are universal and remain unchanged\footnote{The Python implementation of the proposed framework is publicly available at \url{https://github.com/taranO/ml-based_source_localization_from_uv-plane}.}.

\subsection{Input data pre-processing}
\label{sec:preprocessing}

\begin{figure*}[t!]
    \centering
    \begin{subfigure}[c]{.49\textwidth}
      \centering
      \includegraphics[width=\linewidth]{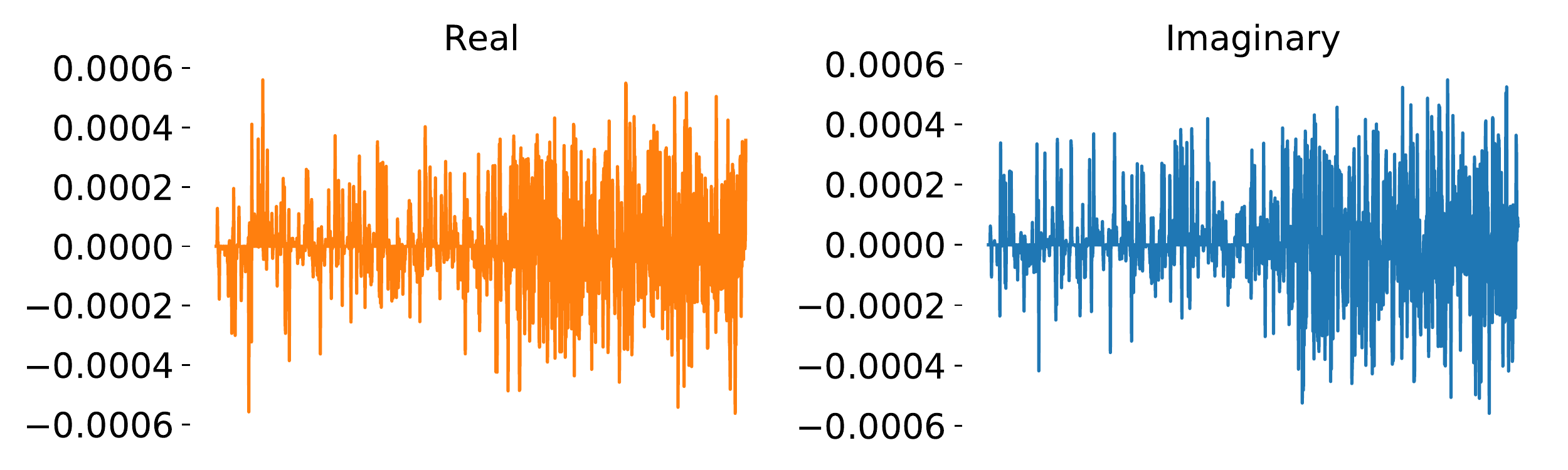}
    \end{subfigure}
    \hfill
    \begin{subfigure}[c]{.49\textwidth}
      \centering
      \includegraphics[width=\linewidth]{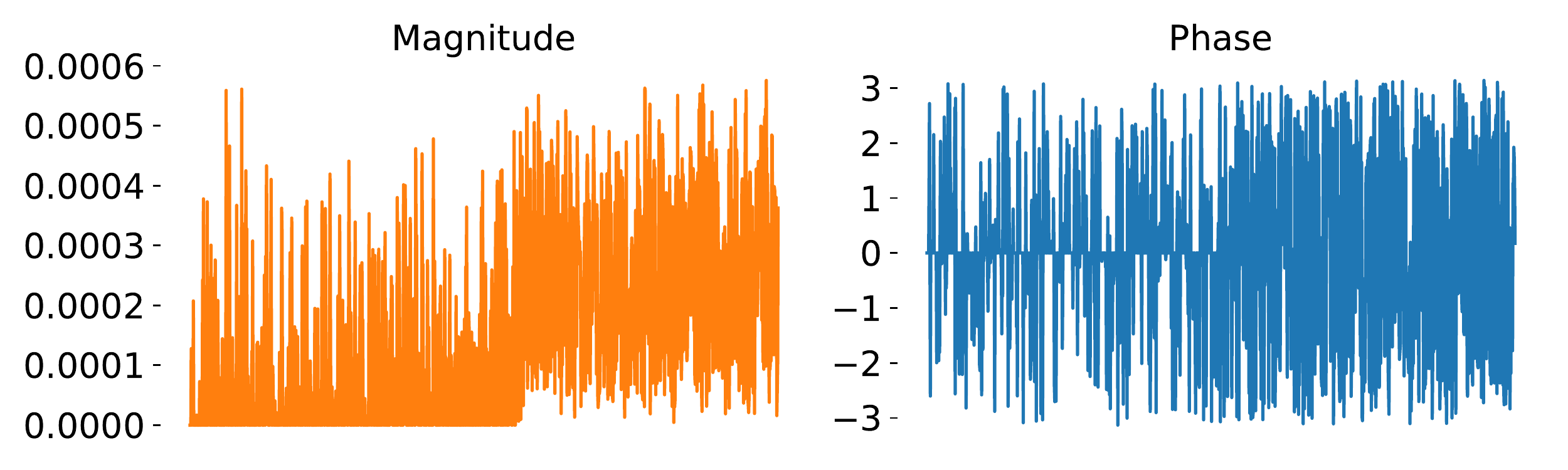}
    \end{subfigure}    
    \begin{subfigure}[c]{.49\textwidth}
      \centering
      \includegraphics[width=\linewidth]{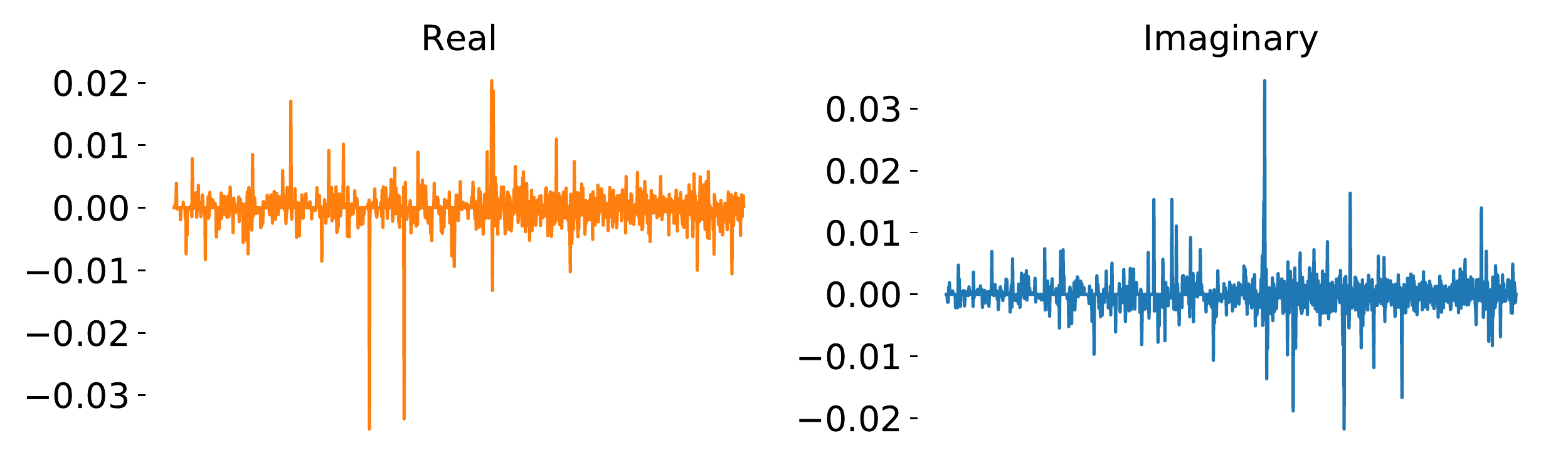}
    \end{subfigure}  
    \hfill
    \begin{subfigure}[c]{.49\textwidth}
      \centering
      \includegraphics[width=\linewidth]{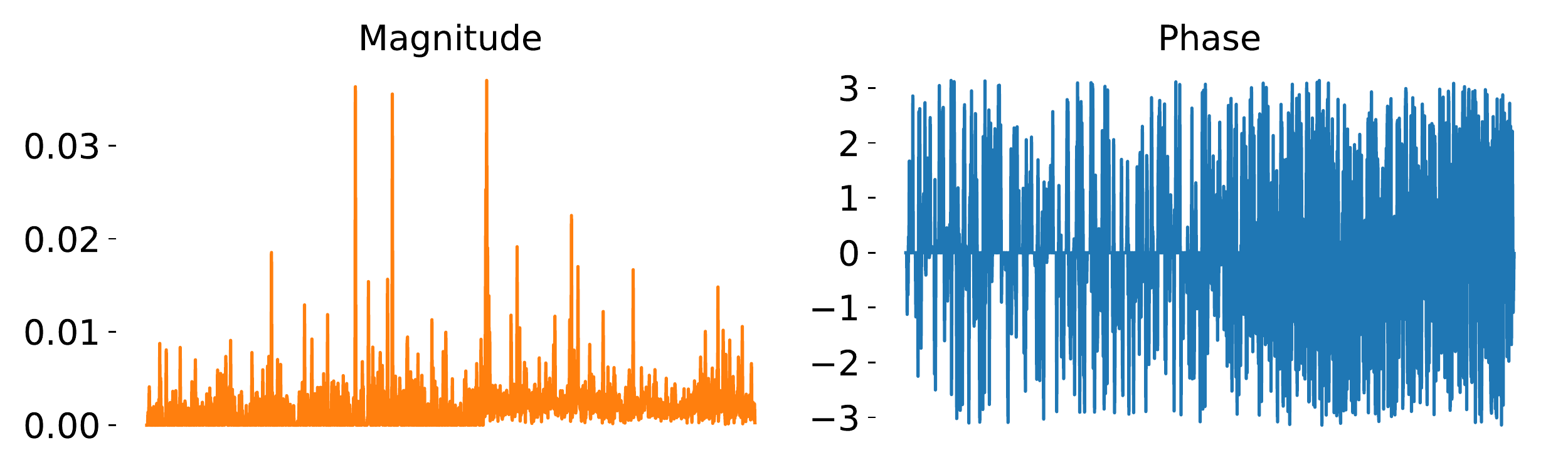}
    \end{subfigure}      
    
    \caption{Examples of input data. Top-left:\ Real and imaginary components in the noise-free case. Top-right:\ Magnitude and phase components in the noise-free case. Bottom-left:\ Real and imaginary components in the noisy case. Bottom-right: Magnitude and phase components in the noisy case.}
    \label{fig:input data example}
\end{figure*}

Examples of the raw noise-free and noisy input data, $\Tilde{\y} \in \mathbb{R}^{2 \times K}$, are given in Fig. \ref{fig:input data example}. As can be seen from the noise-free case, the real and imaginary components have the same dynamic range of power $10^{-4}$ (Fig. \ref{fig:input data example}, top-left), while the magnitude and phase have different dynamic range (Fig. \ref{fig:input data example}, top-right): the magnitude is of a power of $10^{-4}$, similarly to the real and imaginary components, and the phase is in a much wider range, going from $-3$ to $3$. Regarding the noisy case, it is important to note that in the real, imaginary, and magnitude components, the noise dominates and increases the dynamic range of data by two orders of magnitude, while the phase component is less affected by noise and preserves the same dynamic range: from $-3$ to $3$. To guarantee stable DNN training and to avoid the vanishing of gradients, the real and imaginary components are normalized by multiplying by 1000 and clipped in the range $[-10, 10]$. The clipping allows us to reduce the impact of strong outliers. The magnitude component was multiplied by 100 and clipped in the range of $[0, 1]$\footnote{As it can be seen from top-right Fig. \ref{fig:input data example}, the dynamic range of the magnitude is smaller than the dynamic range of the phase component. In this respect, we try to preserve this deviation and use the smaller normalization factor for the magnitude compared to the real and imaginary components.}. The phase component was processed without any normalization. Empirically, this type of normalization was found to be optimal during the proposed model training. The normalized vector is denoted as $\bar{\y} \in \mathbb{R}^{2 \times K}$

\subsection{DNN processing}
\label{subsec:dnn-processing}

It is important to highlight that our goal is to perform the source localization. Thus, we are not interested, for the time being, in the prediction of the sources fluxes and other parameters such as size and so on. In this respect, stages 1 and 2 are trained to minimize the similarity score with respect to the true binary source map, $\m \in \{0, 1\}^{N \times M}$.

\subsubsection{Stage 1: Real and imaginary representation}

\noindent For the real and imaginary representation of the input data the schematic architecture of  the model, $g_\btheta$ is shown in the left panel of Fig. \ref{fig:stage 1 g}.\footnote{The architecture details are given in Table \ref{app tab:stage 1 re im} in the Appendix \ref{apppendix: training of proposed framework}.} At first, the given input low-dimensional representation $\bar{\y} \in \mathbb{R}^{2 \times K}$ is mapped into a higher dimension representation $\a_1 \in \mathbb{R}^{2 \times N\cdot M}$ via a fully connected layer. Then the obtained high dimension representation is reshaped into a square representation $\a_2 \in \mathbb{R}^{2 \times N \times M}$. Finally, the weighted sum of the obtained components produces the output $\hat{\m}_{st.1} \in \mathbb{R}^{N \times M}$. 

\begin{figure*}[t!]
    \centering
    \begin{subfigure}[c]{.195\textwidth}
      \centering
      \includegraphics[width=0.95\linewidth]{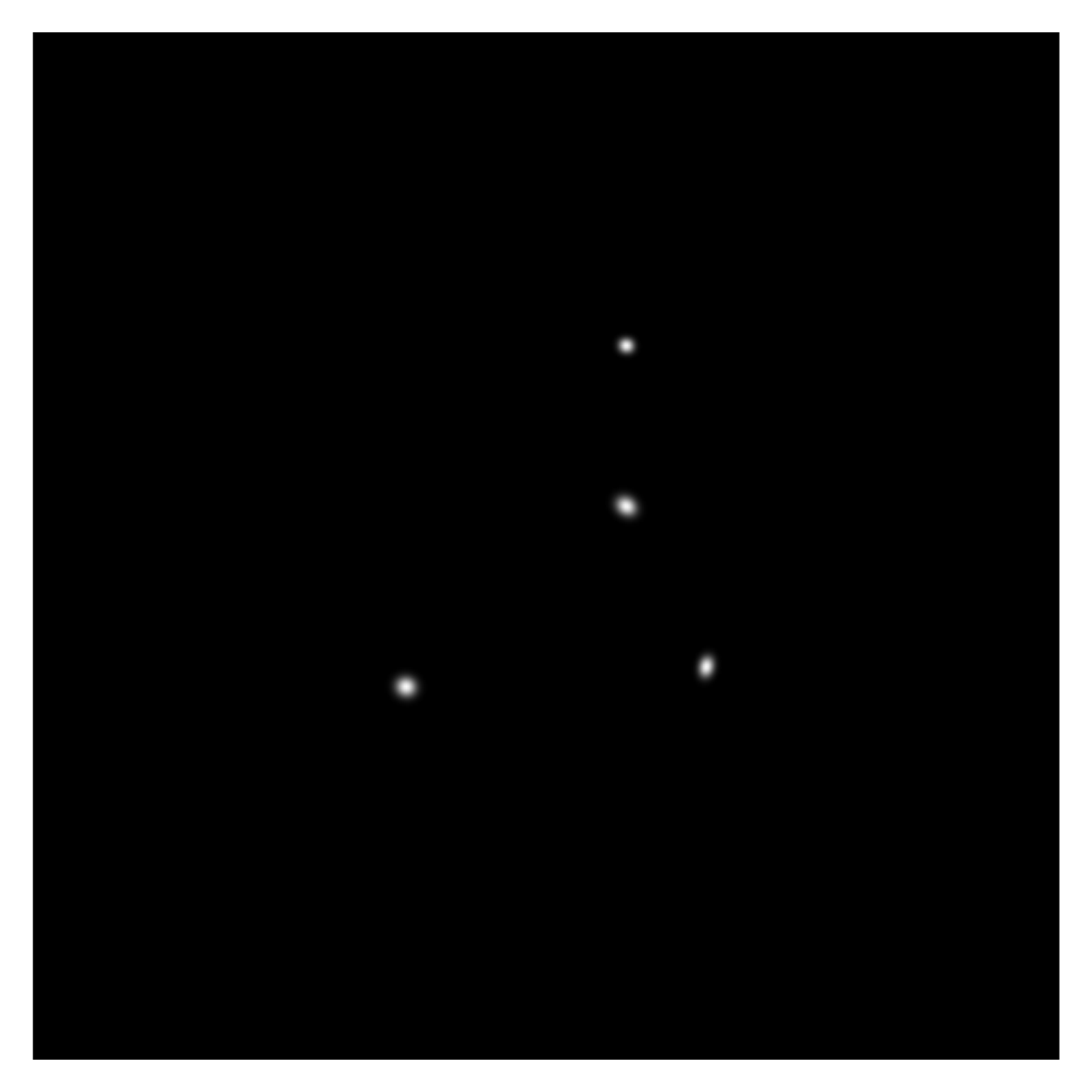}
    \end{subfigure}
    \hfill
    \begin{subfigure}[c]{.195\textwidth}
      \centering
      \includegraphics[width=0.94\linewidth]{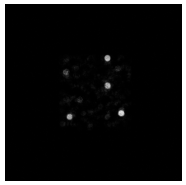}
    \end{subfigure}
    \hfill
    \begin{subfigure}[c]{.195\textwidth}
      \centering
      \includegraphics[width=0.94\linewidth]{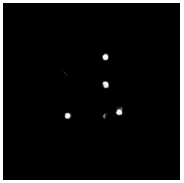}
    \end{subfigure}    
    \hfill
    \begin{subfigure}[c]{.195\textwidth}
      \centering
      \includegraphics[width=0.94\linewidth]{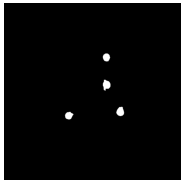}
    \end{subfigure}    
    \caption{Example of the proposed model's outputs. Left to right: Example of the true binary source map $\m \in \{0,1\}^{N \times M}$. Proposed framework estimations, $\hat{\m}_{st.1} \in \mathbb{R}^{N \times M}$ and $\hat{\m}_{st.2} \in \mathbb{R}^{N \times M}$ based on the real and imaginary input data representation (for the magnitude and phase representation the results look similar). Final estimation: $\hat{\m} \in \{0,1\}^{N \times M}$.}
    \label{fig:examples of predictions}
\end{figure*}

\subsubsection{Stage 1: Magnitude and phase representation}

\noindent For the magnitude and phase representation of the input data, the schematic architecture of  the model $g_\btheta$ is shown in right Fig. \ref{fig:stage 1 g}.\footnote{The architecture details are given in Table \ref{app tab:stage 1 mag ph} in the Appendix \ref{apppendix: training of proposed framework}.} First, we performed the weighted element-wise multiplication of the two low-dimensional components in $\bar{\y} \in \mathbb{R}^{2 \times K}$. Then, the resulting low-dimensional representation, $\bb_1 \in \mathbb{R}^{1 \times K}$,  was mapped into a higher dimensional representation, $\bb_2 \in \mathbb{R}^{1 \times N\cdot M}$, via a fully connected layer. Finally, this high-dimensional representation is reshaped to a square representation that corresponds to the output $\hat{\m}_{st.1} \in \mathbb{R}^{N \times M}$ of the stage 1, which is a real valued estimation of the source map. \\

The estimation of the parameters $\btheta$ of the trained model $g_\btheta$ is based on solving the minimization problem:
%
\begin{equation} 
    \centering
    \hat{\btheta} = \argmin_{\btheta} \loss_{st.1}(\btheta) = \loss_{\textrm{mse}}(\m, \hat{\m}_{st.1}),
    \label{eq:optimization problem st1}
\end{equation}
where $\hat{\m}_{st.1} = g_\btheta(\bar{\y})$ and $\loss_{\textrm{mse}}(.\;,.)$ denotes the mean square loss.

\subsubsection{Stage 2}

Due to the fact that the network prediction, $\hat{\m}_{st.1} \in \mathbb{R}^{N \times M}$, is valued as real, the model $g_\btheta$ preserves some information about the true source intensity. Therefore, some predicted sources might be of low intensity, as shown in Fig. \ref{fig:examples of predictions} (the second sub-figure). As a result, after the binarization process, certain sources might be lost, especially in the noisy case. In this respect, stage 2 might be considered as a quality enhancement stage for the source detection where (as can be seen in Fig. \ref{fig:examples of predictions}, i.e., the third sub-figure) the predicted source intensity is close to the binary representation. Taking into account the simple nature of the expected prediction in the form of a simple binary map without any complex shapes and textures, we used a simple auto-encoder model\footnote{The architecture details of the used auto-encoder are given in Table \ref{app tab:stage 2} in the Appendix \ref{apppendix: training of proposed framework}.} as the model
$q_\bphi$. For the more complicated tasks such as  source intensity or any other estimation of physical parameters, the model $q_\bphi$ might be represented by more advanced models, such as UNet \citep{long2015fully} or Transformers \citep{vaswani2017attention}. 

The estimation of the parameters $\bphi$ of the trained model $q_\bphi$ is done by solving the minimization problem:
%
\begin{equation} 
    \centering
    \hat{\bphi} = \argmin_{\bphi} \loss_{st.2}(\bphi) = \loss_{\textrm{mse}}(\m, \hat{\m}_{st.2}),
    \label{eq:optimization problem st2}
\end{equation}
where $\hat{\m}_{st.2} = q_\bphi(g_{\btheta^*}(\bar{\y}))$,  $\btheta^*$ denotes fixed pre-trained model parameters and $\loss_{\textrm{mse}}(.\;,.)$ denotes the mean square loss.

\subsection{Post-binarization and source localization}
\label{subsec:post-binarization and source localization}

Taking into account a need to satisfy the differentiability of DNN for the gradient propagation at the training, the DNN cannot produce the binary outputs. In this respect, the post-binarization stage is necessary. However, instead of performing the hard thresholding that might produce source blobs of an uncontrollable size, a morphological-based binarization was used. 

The morphological binarization consists of determining the optimal threshold value\footnote{More details are given in Appendix \ref{apppendix: training of proposed framework}.} for every particular DNN output, $\hat{\m} \in \mathbb{R}^{N \times M}$, on the fly. After the binarization, the connected neighborhoods are detected, while overly small or big regions are rejected. Finally, the source position was estimated for each detected region by taking their centroids.


\subsection{Conceptual advantages}

Compared to the traditional pipeline, conceptually, the proposed framework has the following advantages. The input data dimensionality is much smaller: only the sampled uv-data are processed to produce a localization binary map, $\m$, while the traditional pipeline works with the full-size uv-plane, where missing frequencies are filled with zeros. Moreover, the proposed framework might be extended to any type of data sources and even for mixed types, while the CLEAN method used in the traditional pipeline offers only a poor reconstruction of regions of extended emission. For this reason, it is usually applied  to the sky models that are only composed of point sources. For regions of extended emission, other methods are used. In this respect, the traditional pipeline is more demanding in terms of the expert knowledge. Finally, the CLEAN reconstruction as well as the dirty image estimation are not required as such to carry out the source detection, which represents a considerable computational advantage in practice.

\begin{figure}[t!]
      \centering
      \includegraphics[width=0.35\linewidth]{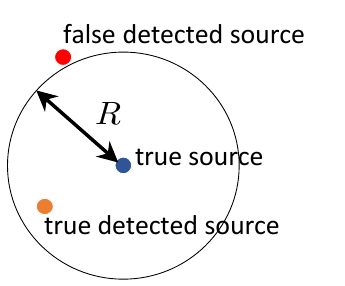}
      \caption{Schematic representation of the acceptable distance of \R{} between the true and predicted sources.}
      \label{fig:detection radius}
\end{figure}

\section{Results and analysis}
\label{sec:results and analysis}

By default and unless specified, all results are given as an average over three test subsets mentioned in Sect. \ref{sec:data}.
For simplicity, we named the proposed framework based on the real and imaginary components of the input data representation as "\reim"{} and those based on the magnitude and phase components are given as "\magphas{}."

\subsection{Metrics}
\label{sec:metrics}

We cross-matched detected sources with true sources in each sky model using a match radius, $R,$ equal to the beam size as shown in Fig. \ref{fig:detection radius} to compute the model performance metrics listed below. The impact of the matched radius in the results is discussed in Sect. \ref{sec:results on noisy data}.

To evaluate the performance of the traditional pipeline and the proposed framework, we used the purity and completeness metrics that are defined as follows.

Purity shows the fraction of the true sources among all detected sources:
\begin{equation}
    \textrm{Purity} = \frac{\textrm{TP}}{\textrm{TP} + \textrm{FP}},
    \label{eq:purity}
\end{equation}
where $\textrm{TP}$ and $\textrm{FP}$ denote true positive and false positive sources, respectively. Here, $1 -\textrm{purity}$ represents the fraction of false detections.

Completeness is equal to the fraction of true sources that are successfully detected:
\begin{equation}
    \textrm{Completeness} = \frac{\textrm{TP}}{\textrm{TP} + \textrm{FN}}
    \label{eq:completeness}
\end{equation}
where $\textrm{FN}$ represents the false negative (missing) sources.

%
\subsection{Results on noise-free data}
\label{sec:results on noiselsess data}

We consider the noise-free case as an ideal condition measurement. In real observations, this reflects a hypothetical situation, since all measurements are corrupted by noise of different nature. However, we found it to be important to validate the performance of the investigated approaches under the assumption that there is no noise to understand their baseline performance.  

\begin{figure}[t!]
      \centering
      \includegraphics[width=1\linewidth]{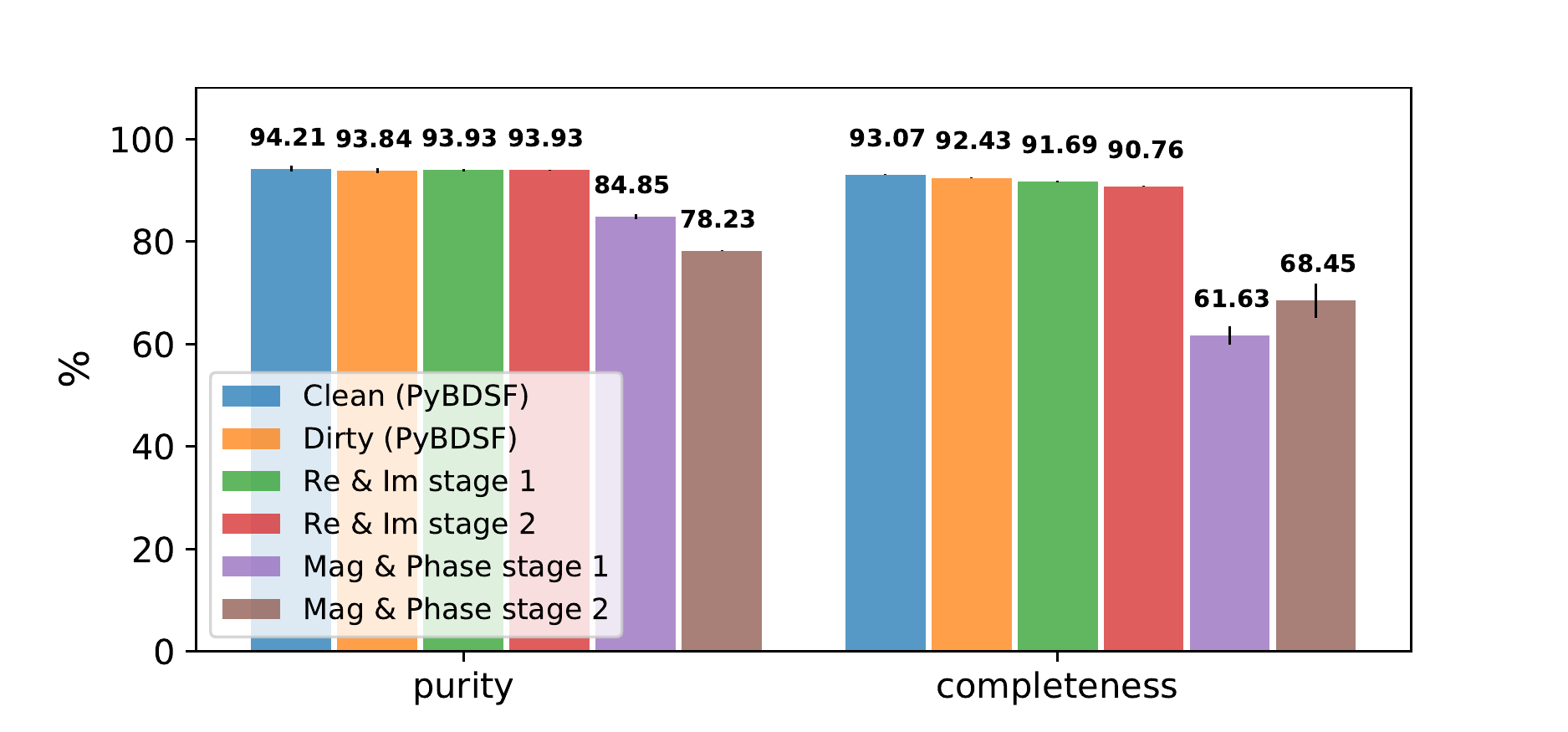}
      \caption{Purity and completeness of the methods under investigation in the noise-free case. Small black lines on the top of each bar correspond to the standard deviation from the average.}
      \label{fig:bar noise-free}
\end{figure}

\begin{figure}[t!]
    \centering
    \begin{subfigure}[c]{.49\linewidth}
      \centering
      \includegraphics[width=0.8\linewidth]{images/proposed_framework/predicted_noiseless_st1.png}
    \end{subfigure}
    \hfill
    \begin{subfigure}[c]{.49\linewidth}
      \centering
      \includegraphics[width=0.8\linewidth]{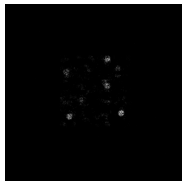}
    \end{subfigure}
    \caption{Difference between the noise-free (left) and noisy (right) prediction of the proposed \reim{} stage 1 for a simulated sky model.}
    \label{fig:noise-free vs noisy re im st1}
\end{figure}
Figure \ref{fig:bar noise-free} demonstrates the purity and completeness obtained for PyBDSF applied to the CLEAN and dirty images and the proposed framework in four different configurations. We chose the source detection parameters in such a way as to have the maximum completeness under the assumption that the acceptable purity should be about 94 \%\footnote{The used \pybdsf{} parameters are \textit{thresh\_pix} = 7 and \textit{thresh\_isl} = 5. In the proposed framework \textit{area\_lim} parameter was set to 125 in case of \reim{} at stage 1 and to 200 at stage 2. In \magphas{} the same parameter was set to 300 and 310 at stage 1 and stage 2 correspondingly.}. It is easy to see that \pybdsf{} on the CLEAN and dirty images and the proposed \reim{} framework exhibit a similar performance. 
There is no big difference in the performance of \pybdsf{} for the CLEAN and dirty images. The \reim{} stage 1 and 2 provide very close results. In the case of \magphas{}, the performance is worse. As it is mentioned in Sect. \ref{sec:preprocessing}, there is a big difference in the dynamic range of the magnitude and phase components. On one side, such a difference is natural and should be preserved. On the other side, it is a disadvantage and a challenge for the DNN training. This explains the obtained non-optimal results. 

\begin{figure}[t!]
      \centering
      \includegraphics[width=1\linewidth]{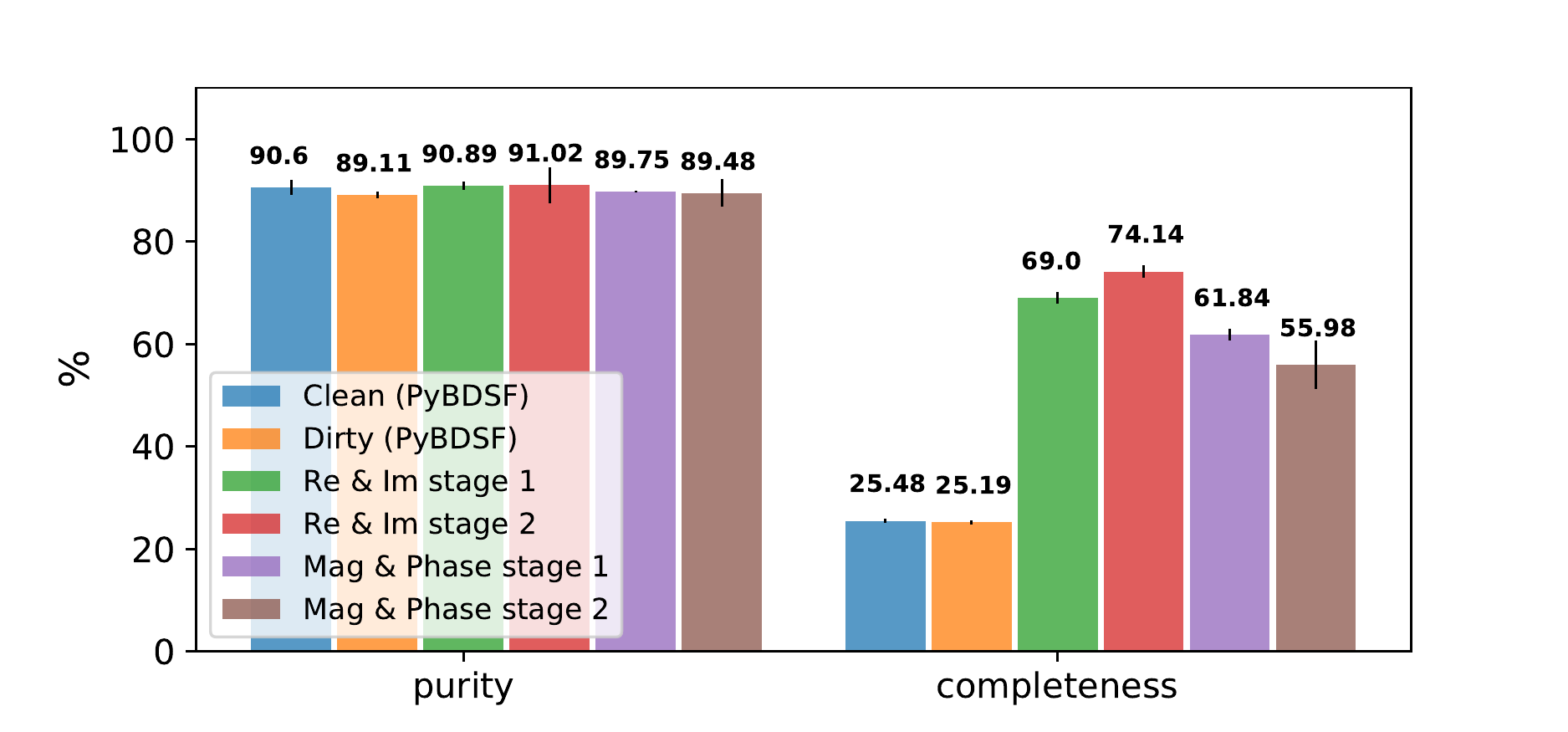}
      \caption{Purity and completeness of the methods under investigation in the noisy case, averaged over all S/N values. The small black lines on the top of each bar correspond to the standard deviation from the average.}
      \label{fig:bar noisy}
\end{figure}

\subsection{Results on noisy data}
\label{sec:results on noisy data}

The noisy scenario is of a particular interest for our study of real observations. In this respect, we provide a more detailed analysis of the obtained results.

Before presenting the results of the proposed framework for the noisy case, we would like to underline the difference in the predictions between the noise-free and noisy case. It helps gain an understanding of why the proposed framework is better than the traditional pipeline. The noise-free and noisy predictions of the \reim{} after stage 1 are shown in Fig. \ref{fig:noise-free vs noisy re im st1}. It is important to highlight that in contrast to the CLEAN or dirty noisy data used in the traditional pipeline (Fig. \ref{fig:simulation examples}), the proposed framework predictions based on the noisy input data are free from the background noise, which is very important for the efficient source localization.

In Fig. \ref{fig:bar noisy}, we show the purity and completeness for \pybdsf{} and the proposed framework in four different configurations. For a fair comparison, the parameters of source localization for all methods under investigation are selected in such a way to have a purity of about 90 \%\footnote{We set \pybdsf{} parameter \textit{rms} to 4.2450E-05 for the CLEAN data and to 4.275E-05 for the dirty case. In the proposed framework \textit{area\_lim} parameter was set to 75 in case of \reim{} at stage 1 and to 240 at stage 2. In \magphas{} the same parameter was set to 200 and 340 at stage 1 and stage 2 correspondingly.}. We fixed and used these parameters for all the following experiments.

In comparison to the noise-free case,  in the noisy case we have slightly smaller purity (about 3-4 \% less), but the obtained completeness is significantly smaller for all methods under investigation. Secondly, it is important to note that the \pybdsf{} completeness does not exceed 25 - 26 \%, which is very small for practice. At the same time, it is interesting to note that the proposed \magphas{} framework that has the worst performance in the noise-free case does, in fact, outperform the \sota{} \pybdsf{}  on the noisy data. Its completeness is about 60 \% after stage 1 and about 55 \% after stage 2. The proposed \reim{} framework demonstrates the best completeness about 69 \% after stage 1 and 74 \% after stage 2. However, it should be pointed out that stage 1 is more stable, as shown by the lower standard deviation. 

\begin{figure}[t!]
      \centering
      \includegraphics[width=0.9\linewidth]{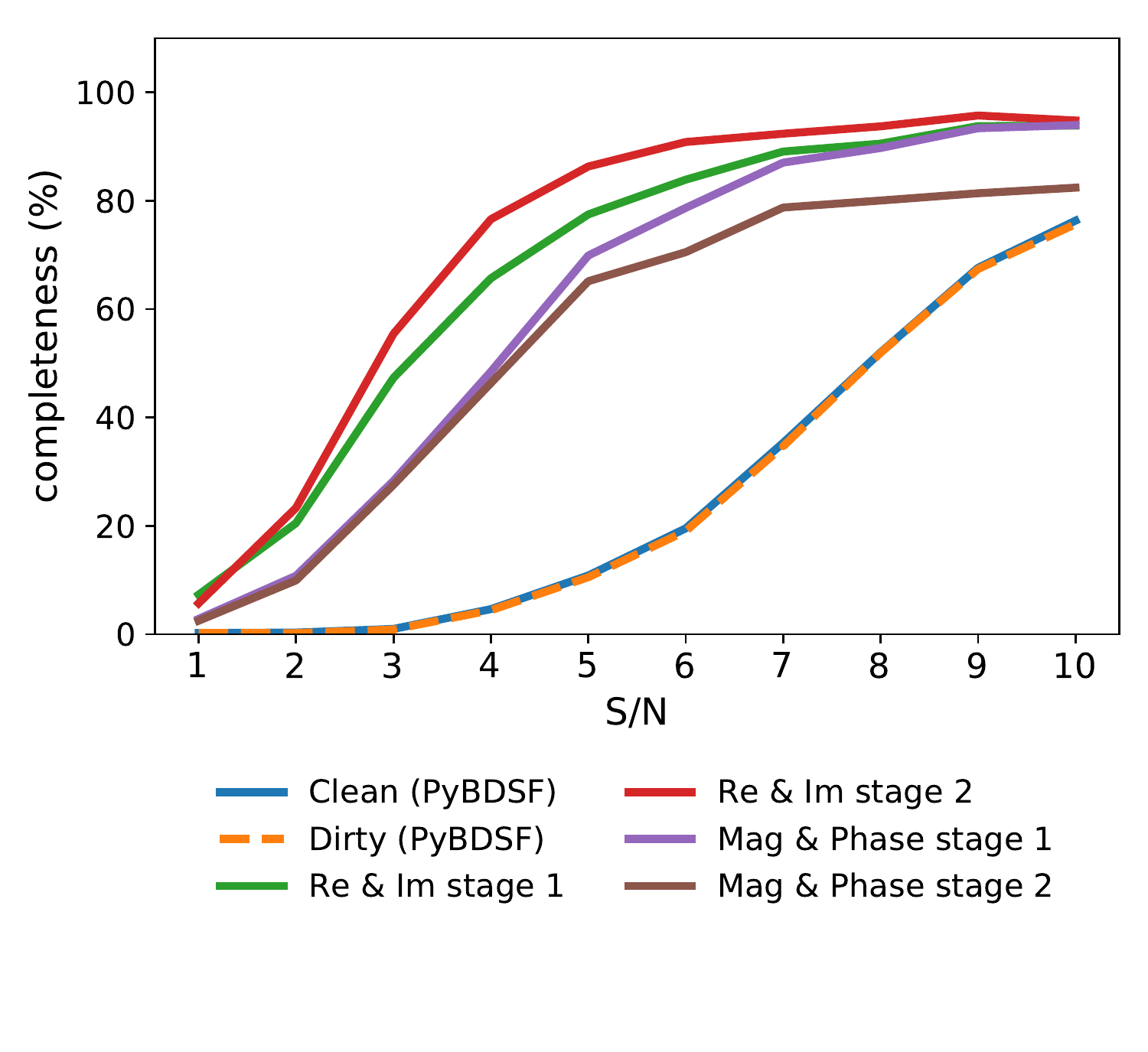}
      \vspace{-0.75cm}
      \caption{Dependence of completeness on S/N of the sources.}
      \label{fig:snr sources statistic}
\end{figure}

\begin{figure*}[t!]
    \centering
    \begin{subfigure}[c]{.32\textwidth}
      \centering
      \includegraphics[width=0.95\linewidth]{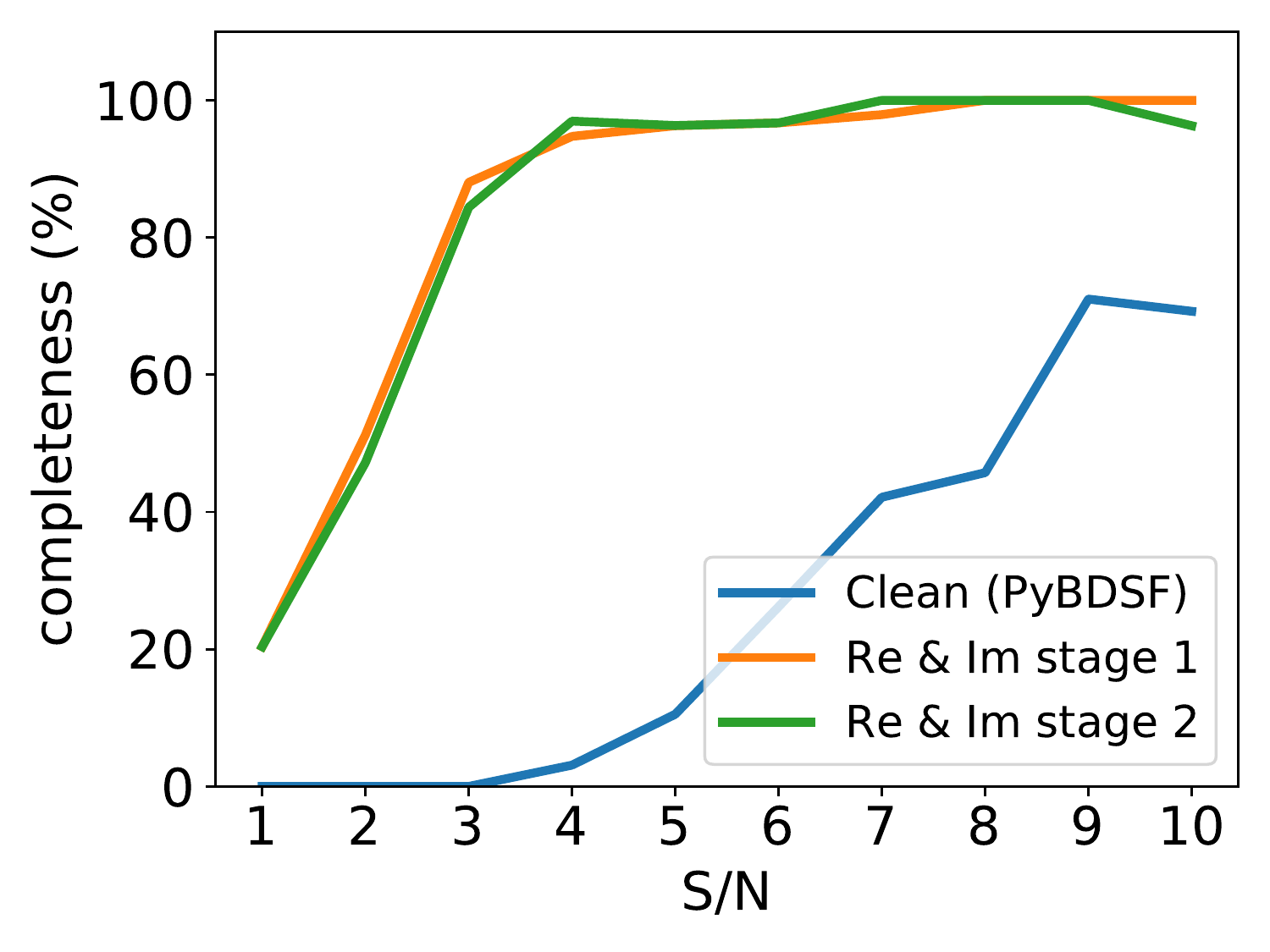}
    \end{subfigure}
    \hfill
    \begin{subfigure}[c]{.32\textwidth}
      \centering
      \includegraphics[width=0.95\linewidth]{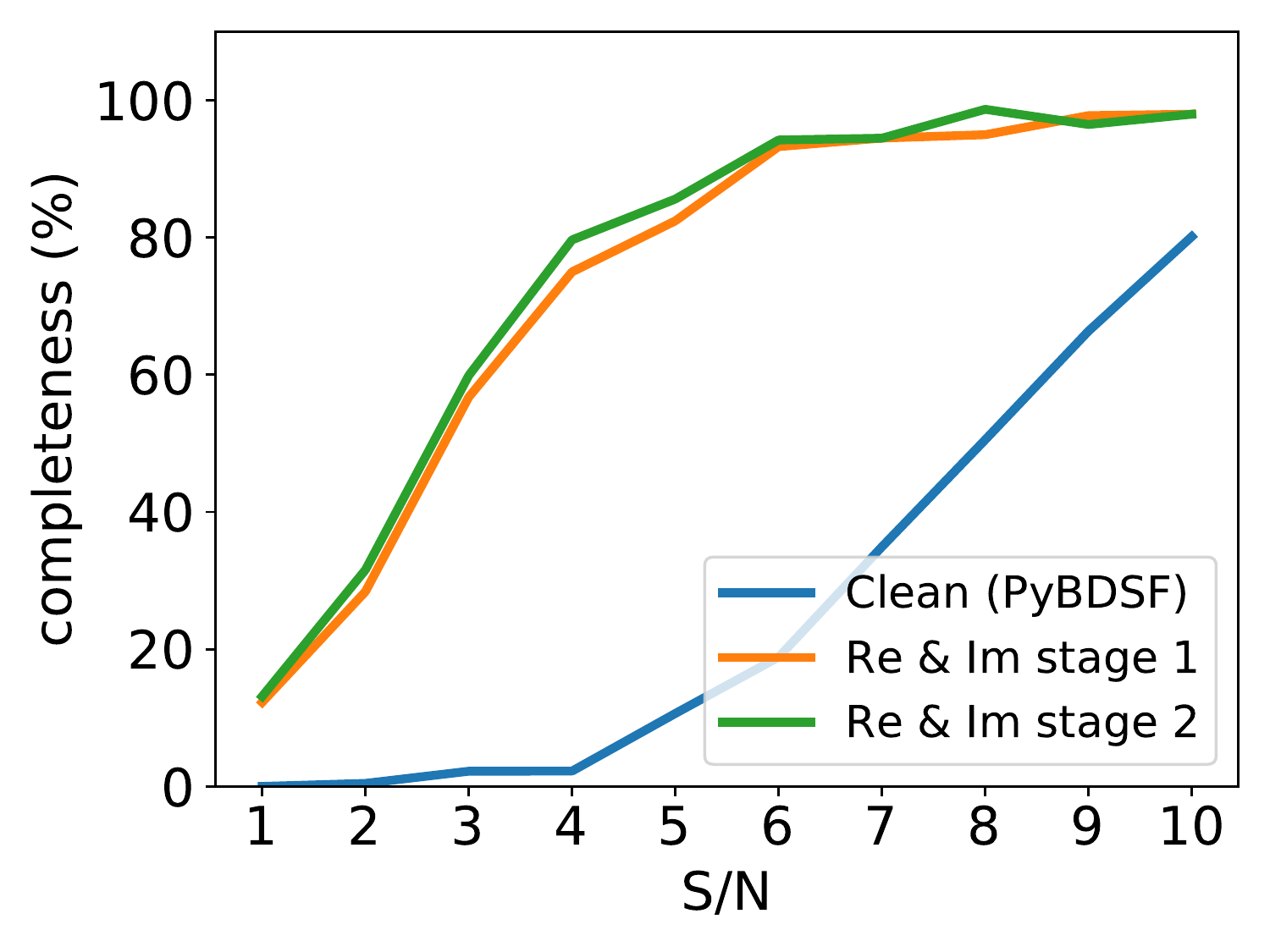}
    \end{subfigure}    
    \hfill
    \begin{subfigure}[c]{.32\textwidth}
      \centering
      \includegraphics[width=0.95\linewidth]{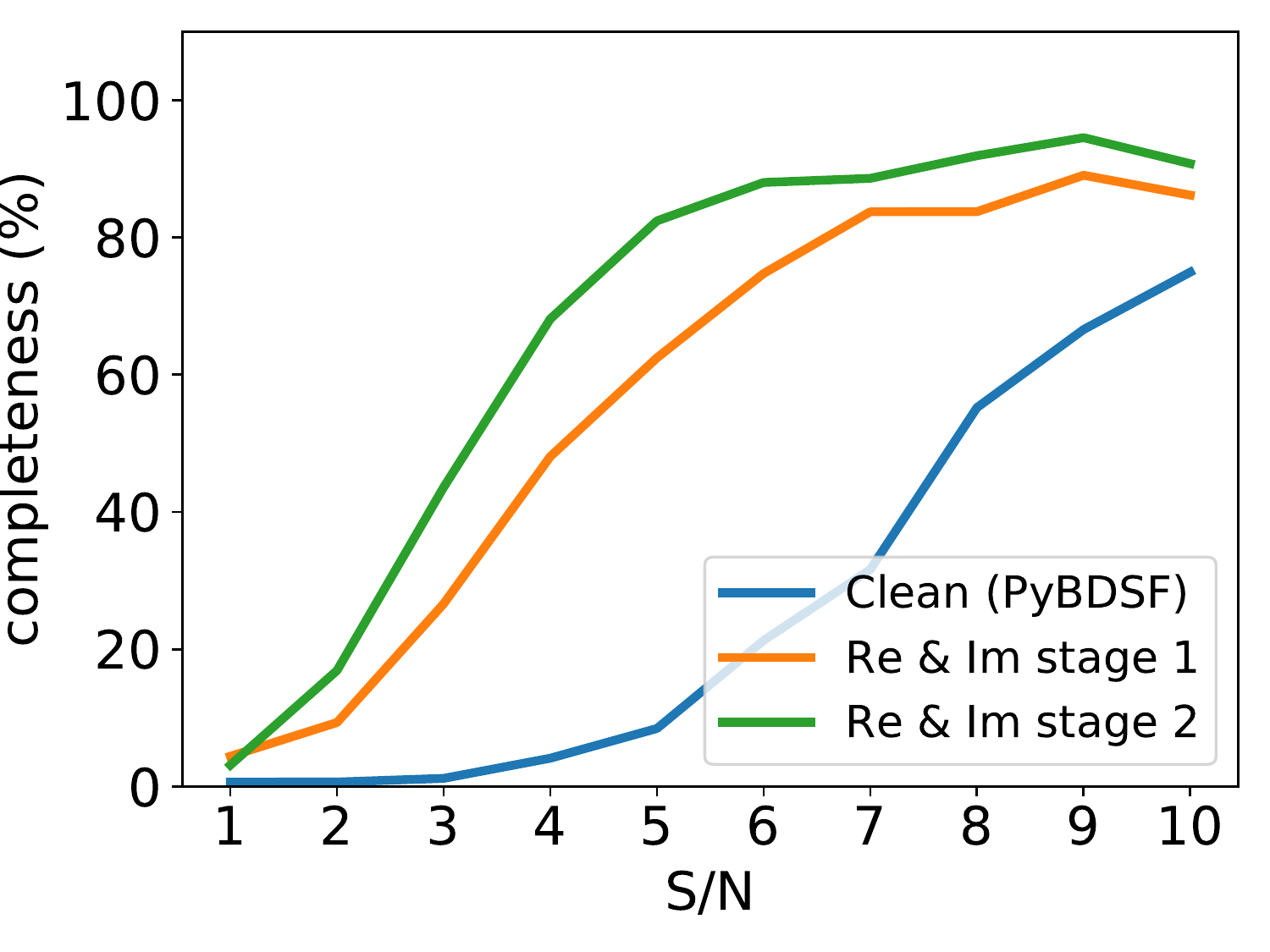}
    \end{subfigure}  
    \caption{Dependence on the efficiency of \pybdsf{} and the proposed \reim{} framework on the number of sources in the sky model. Left: one source. Middle: three sources. Right: five sources.}
    \label{fig:snr performance by sources}
\end{figure*}

\begin{figure}[t!]
    \centering
    \includegraphics[width=0.9\linewidth]{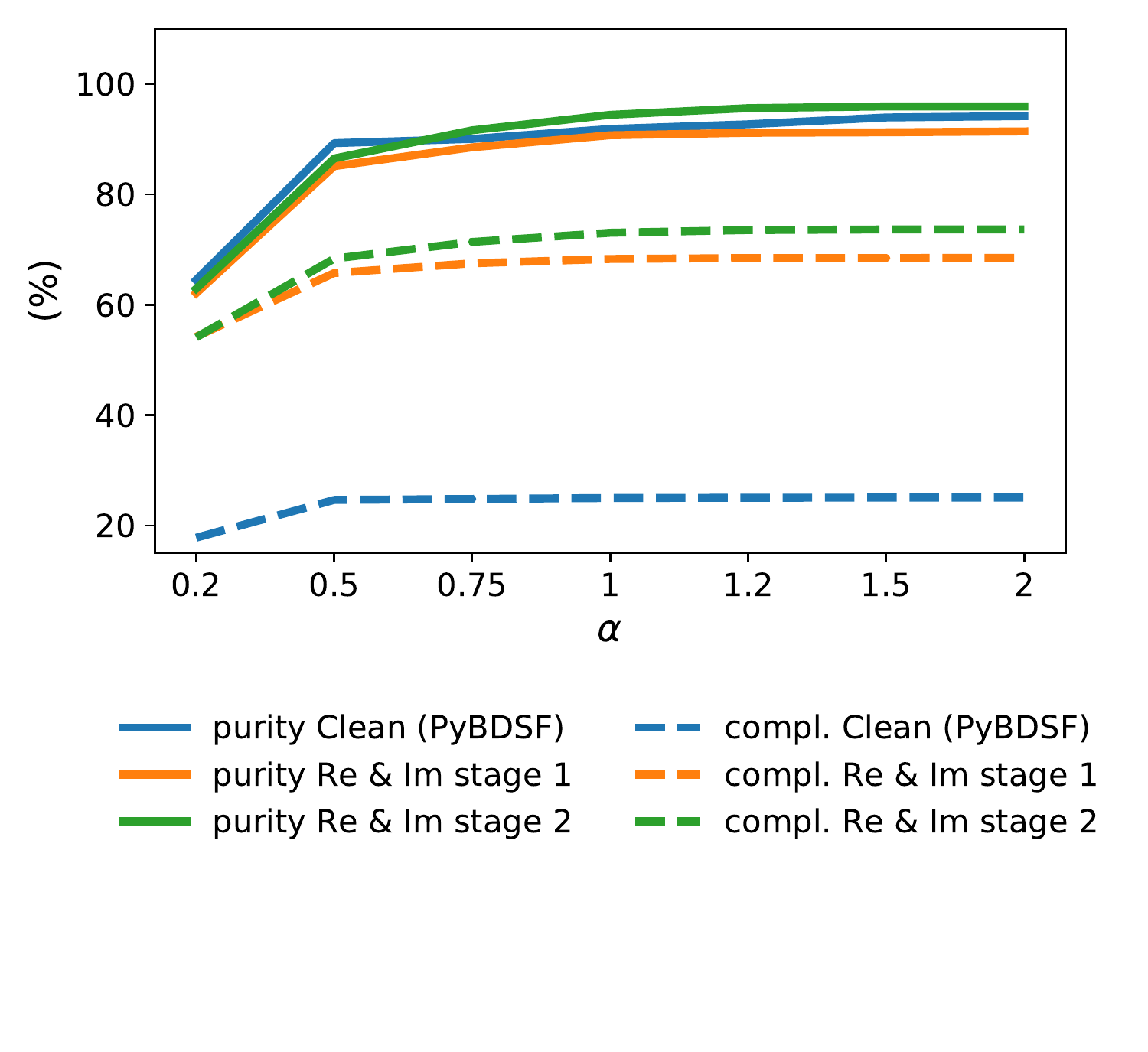}
    \vspace{-1cm}
    
\caption{Dependence on $\alpha$, where $\alpha = \frac{R}{\textrm{beam size}}$ and $R$ is the distance between the true and predicted sources, in the noisy case, averaged over all S/N values.}
    \label{fig:efficiency wrt r}
\end{figure}

\begin{figure}[t!]
    \centering
    \includegraphics[width=1\linewidth]{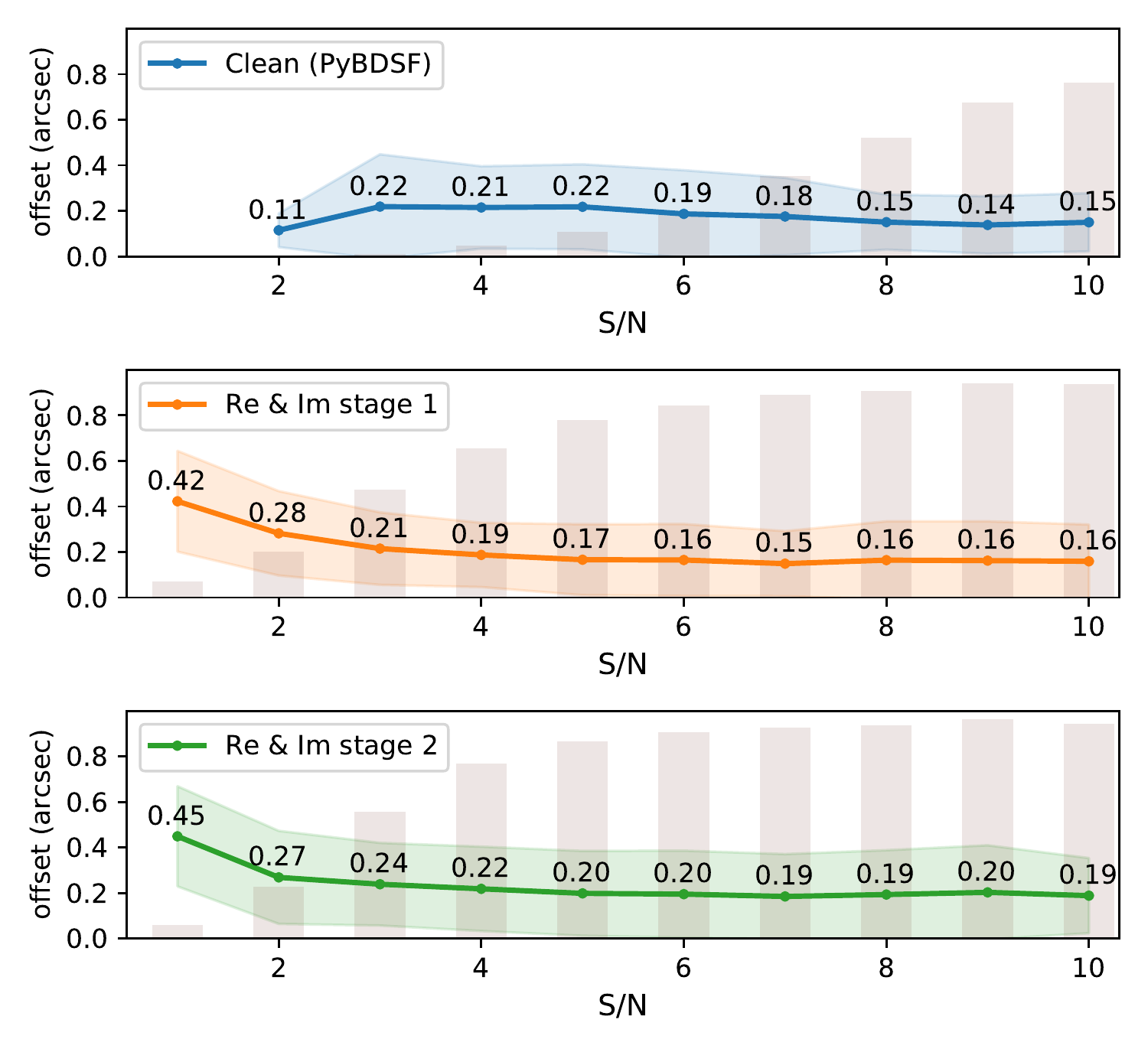}
    \caption{Mean distance (the y-axis) between the true and predicted sources in the noisy case. The coloured semi-transparent background shows the standard deviation from the mean value. The gray bars show the relative number of detected sources.}
    \label{fig:dist mmean std}
\end{figure}

It is obvious that the performance of the source localization highly depends on the S/N value. In Fig. \ref{fig:snr sources statistic}, we plot the completeness obtained for each method under investigation with respect to the S/N of the source. 
We can see that when S/N $\le 3$ \pybdsf{} cannot detect many sources (the completeness does not exceed 2-3 \%). Then, for S/N $\sim$ 2, the proposed \magphas{}{} reaches 10 \% and for S/N $\sim$ 3 it already has a 25 \% completeness. The performance of the proposed \reim{} is even better: for S/N $\sim$ 2 the completeness is about  20 \% and for S/N $\sim$ 3 it is about 50 \% after stage 1 and about 55 \% after stage 2. With the increasing S/N, the performance of all methods increases. However, for the maximum S/N the \sota{} \pybdsf{} can get the completeness only up to about 75 \%, while the proposed \reim{} achieves 93 - 94 \%. It is also important to note that the \reim{} stage 2 offers improvement for S/Ns of 2 - 5, as compared to the \reim{} stage 1. For the very low or high S/N, the performance at both stages is similar.

As mentioned in Sect. \ref{sec:data}, one of the challenges of our data set is closely located sources that are difficult to distinguish, especially in the noisy case with increasing source densities. It is obvious that when we have more sources in the sky model, the probability to have close sources is higher. We analyze the performance of \pybdsf{} and the proposed \reim{} with respect to the source S/N specifically for different numbers of sources in the sky models. The derived results are shown in Fig. \ref{fig:snr performance by sources}. In general, the obtained behavior is similar to that shown in Fig. \ref{fig:snr sources statistic}. However, there are several interesting facts that we can take note of. First of all, the behavior of \pybdsf{} is stable and does not depend too much on the number of sources in the sky model. On the other hand, when the number of sources increases the efficiency of the proposed framework decreases, especially for the low S/N values. For example, for S/N $\sim$ 2 in the case with only one source (Fig. \ref{fig:snr performance by sources} the left panel), the completeness is about 50 \%, while in the case with five sources (Fig. \ref{fig:snr performance by sources} the right panel), the completeness drops down to about 15 - 20 \%. With the S/N increase, the drop in efficiency slows down. This phenomenon is naturally connected with the increased entropy of multi-source spectra. Keeping the fixed number of samples in the uv-plane, the sky model with several sources is characterized by a lower entropy spectrum. Therefore, there is a fundamental accuracy limit of source detection under the restricted sparse sampling of the uv-plane. This is one of the directions for our future investigation. At the same time, it is important to point out that for the setup under investigation and the chosen antenna configuration (for mode details, see Sect. \ref{sec:data}), the expected number of sources ranges from one to five. For the larger number of sources, the ambiguity in the under-sampled uv-data increases. For this reason, the configuration with the larger number of antennas should be used. Under the chosen setup, even though the completeness of \reim{} drops, its general efficiency is much better than \pybdsf{}.

Another important question that we study is the dependency of source localization accuracy on the distance to the true sources. In Fig. \ref{fig:detection radius}, we schematically show the locations of the true source and the corresponding predicted source. We consider the predicted source as true if it is inside a given radius, \R,{} around the true source. In the current experiments, we set an $\R = \alpha \cdot \textrm{beam size}$\footnote{The size of the beam used in the simulation is about 0.82\arcsec.} where $\alpha$ is a scale factor. The results shown in Fig. \ref{fig:efficiency wrt r} are obtained for the test subset 1\footnote{In Fig. \ref{fig:bar noisy}, we can see significant deviation from the average for \reim{} stage 2 in comparison to \reim{} stage 1. This explains why  the performance of stage 2 is better than that of stage 1 in Fig. \ref{fig:efficiency wrt r}.}. It is important to mention that in all previous experiments $\alpha = 1$ or, in other words, the radius $\R$ equals to the beam size. In Fig. \ref{fig:efficiency wrt r}, we can observe some increase in the purity when $\alpha > 1$, but the saturation is achieved already for $\alpha = 1.2 - 1.5$. For the completeness, the saturation is achieved for $\alpha = 1$. 
In Fig. \ref{fig:dist mmean std}, we plot the mean distance in arcsec between the centers of true and detected sources. The gray bars show the relative number of detected sources. The relative means were divided by the total number of true sources for the given S/N. It should be pointed out that in the case of the \reim{} framework, the deviation from the mean value is quite stable and does not depend a great deal on the number of detected sources or S/N. At the same time, with the increase in the S/N and the number of detected sources, the mean value convergences to about 0.16\arcsec after 
 stage 1 and to about 0.19\arcsec after stage 2. For \pybdsf, we can see quite small deviation from the mean values for S/N = 2 and a large deviation for S/N = 3. This can be explained by the very small number of detected sources and, as a consequence, the poor statistics. With the increase in the S/N and the number of detected sources, the deviation from the mean value decreases and the mean value convergences to about 0.15\arcsec. It is important to note that \reim{} framework reaches the convergence at S/N = 5,  while for \pybdsf{} we can observe the convergence only after S/N = 8.

\begin{table*}[t!]
        \centering
        \renewcommand*{\arraystretch}{1.5}
        \caption{Execution time for the source localisation per sky model (noisy data).}  
        \begin{tabular}{lccc} \hline
                \multirow{2}{*}{Approach} & \multicolumn{2}{c}{Execution CPU time, sec}  \\ 
                & reconstruction from uv & source localization & total \\ \hline
                dirty + CLEAN + \pybdsf{}& 17.29 & 1.48 & 18.77 \\
                dirty + \pybdsf{}        & 13.97 & 1.51 & 15.48 \\ 
                \reim{} stage 1          & 0.09  & 0.18   & 0.27 \\ 
                \reim{} stage 1 and 2    & 0.15  & 0.34 & 0.49 \\
                \magphas{}{} stage 1     & 0.25  & 0.22 & 0.47 \\ 
                \magphas{} stage 1 and 2 & 0.33  & 0.51 & 0.84 \\ \hline 
        \end{tabular}
        \label{tab:exec time}   
\end{table*}    

\subsection{Execution time}
\label{sec: execution time}

To investigate the question of the time complexity, we measured the execution time of the source localization for the methods under investigation. The obtained CPU time in sec\footnote{The inference of the proposed framework might be efficiently run at GPU. However, taking into account that the traditional methods are usually run on CPU and  that not all users have access to GPU, we performed the comparison with respect to CPU to have a fair comparison and to show that the proposed framework is easily runnable on CPU.} is summarized in Table \ref{tab:exec time}. For the proposed framework, the column labeled "reconstruction from uv" corresponds to the DNN-processing, Sect. \ref{subsec:dnn-processing}. The  "source localization" column corresponds to the procedure explained in Sect. \ref{subsec:post-binarization and source localization}. It should be pointed out that the recovery of the dirty image takes about 14 sec. The recovery of the CLEAN image takes about 17 sec. On the other hand, the proposed framework in the slowest case (\magphas{} after two stages) estimates the real-valued source map only in 0.33 sec. The source localization by \pybdsf,{} compared to the most efficient \reim{} stage 2 is about 4.3 times slower. In terms of the total time complexity, \pybdsf{} on the CLEAN data (the best traditional pipeline results) is $\sim$ 38 times slower than the proposed \reim{} stage 2 (the best results) -- even on the CPU execution. 

\section{Discussion}
\label{sec:discussion}

\subsection{Performance comparisons with the literature}

In classical approaches to interferometric observations, fidelity and completeness are two important measures of the significance and ``statistical importance'' resulting from source detections and characterizations, which are generally determined by empirical methods. For example, the fidelity is computed by comparing the number of sources detected with positive flux to those in the negative image ($\textrm{Fidelity} = 1 - \frac{N_{neg}}{N_{pos}}$) to empirically determine a S/N threshold above which individual sources are considered "reliably detected'' \citep[see, e.g.,][]{Aravena2016The-ALMA-Spectr,bethermin2020alpine}.
Furthermore, completeness is generally determined using Monte Carlo source injections in the observational data (e.g., in the image plane).

Since fidelity requires a flux measurement (which is not included in the proposed framework), it is not possible to compare  the performance of our method for this quantity, as already mentioned above (Sect. \ref{sec:metrics}). However, we have compared our results to those of \citet{bethermin2020alpine}, who presented a detailed analysis of the completeness from extra-galactic ALMA continuum  observations with properties similar to those of our sky simulations. Following \citet{bethermin2020alpine}, we therefore determined the completeness as a function of the normalized injected flux: 
    \begin{equation}
        f_{\rm norm} = \frac{\textrm{total flux}}{\sigma_{\textrm{noise}}} \cdot \frac{b_{min}\cdot b_{maj}} {\sqrt{b_{min}^2 + s_{min}^2} \cdot \sqrt{b_{maj}^2 + s_{maj}^2}},
        \label{eq:snr2}
    \end{equation}  
    
where $b_{min}$ and $b_{maj}$ correspond to the minor and major axes of the beam, and $s_{min}$ and $s_{maj}$ denote the minor and major axes of the source. This quantity resembles an effective, normalized S/N, and it encapsulates in particular variations of sources sizes and provides a simple functional description of the completeness, as shown by \citet{bethermin2020alpine}.

\begin{figure}[t!]
      \centering
      \includegraphics[width=0.9\linewidth]{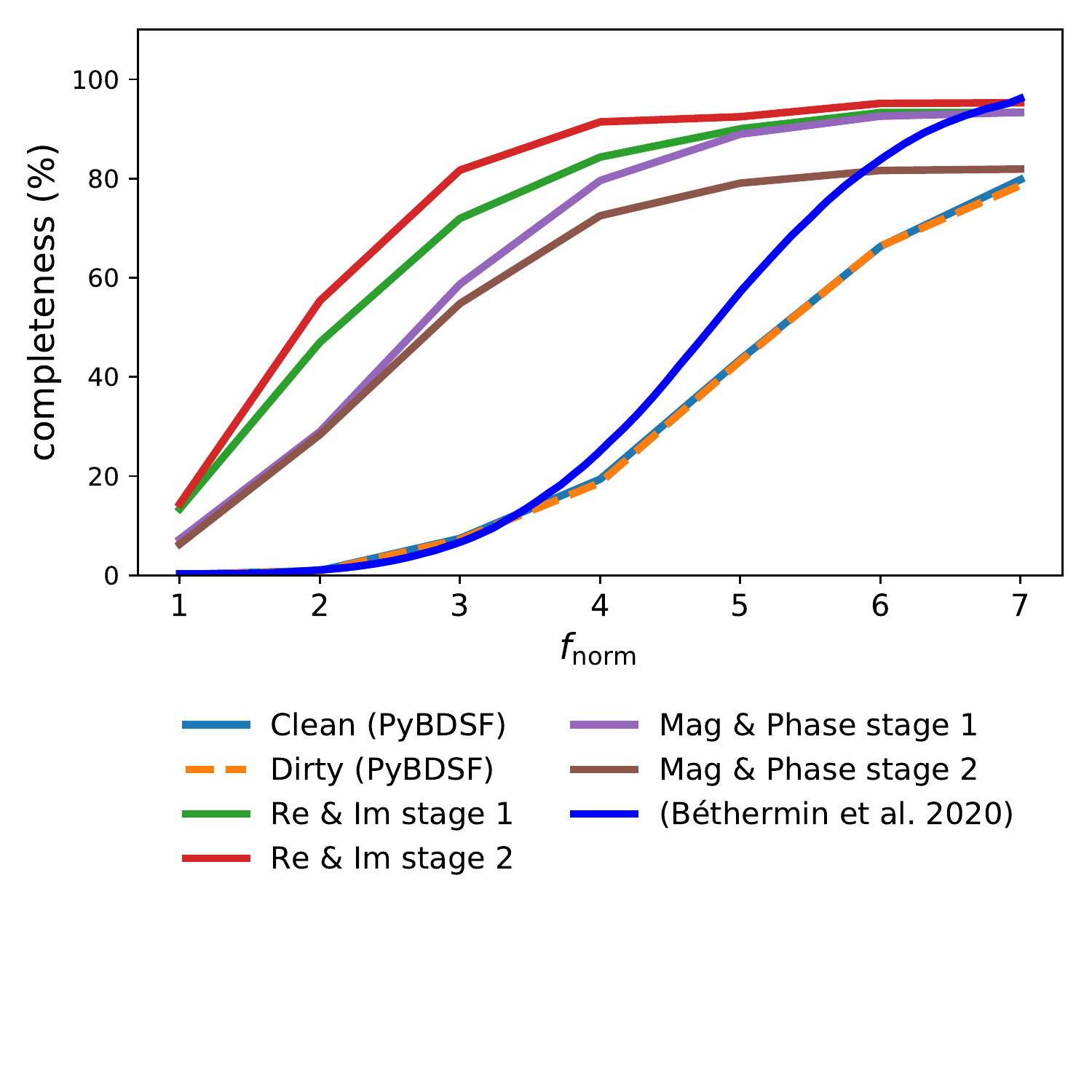}
      \vspace{-1.5cm}
      \caption{Dependence of completeness on normalized S/N compared to \citet{bethermin2020alpine}.}
      \label{fig:snr2 results}
\end{figure}

\begin{figure*}[t!]
    \centering
    
    \begin{subfigure}[c]{.32\linewidth}
      \centering
      \includegraphics[width=1\linewidth]{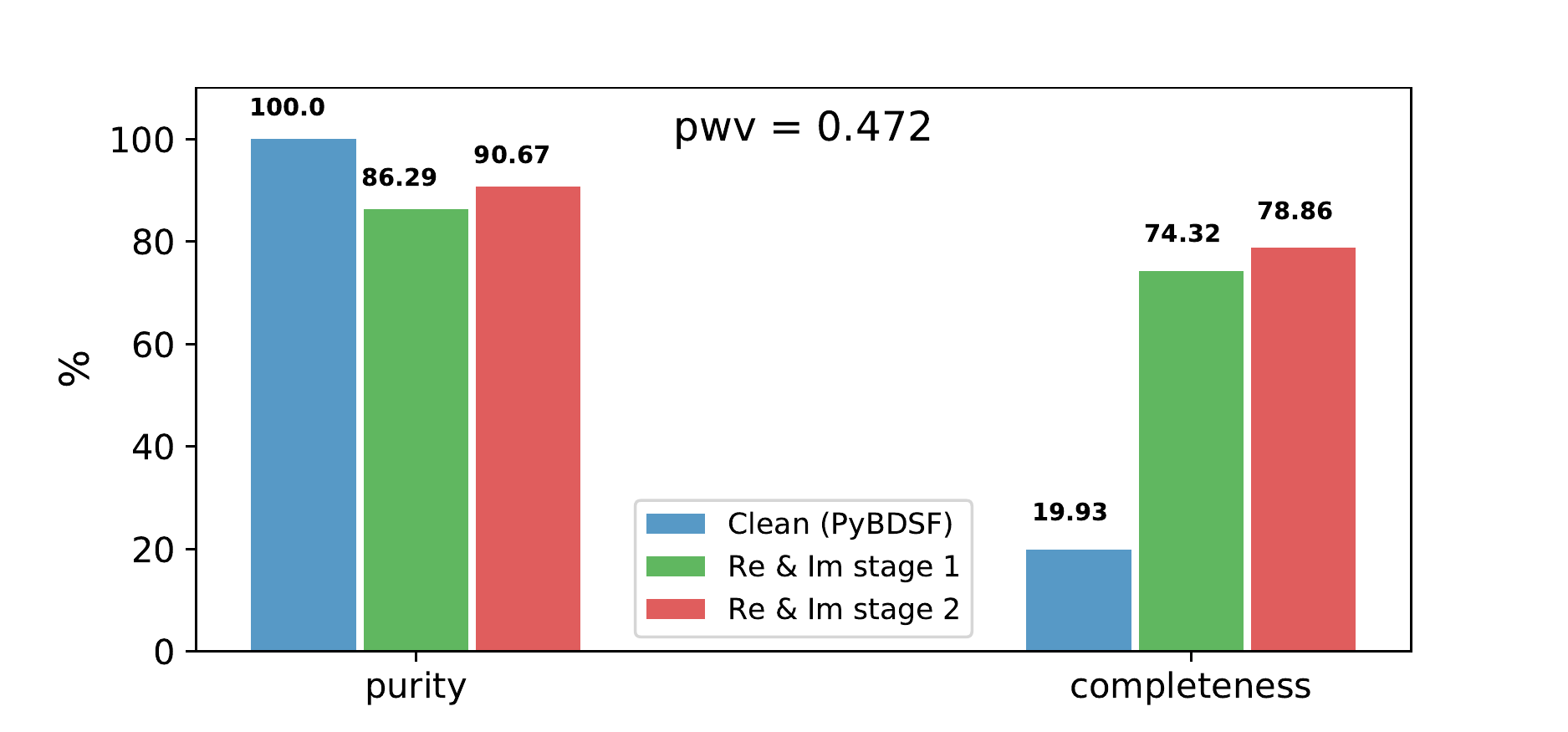}
    \end{subfigure}
    \hfill
    \begin{subfigure}[c]{.32\linewidth}
      \centering
      \includegraphics[width=1\linewidth]{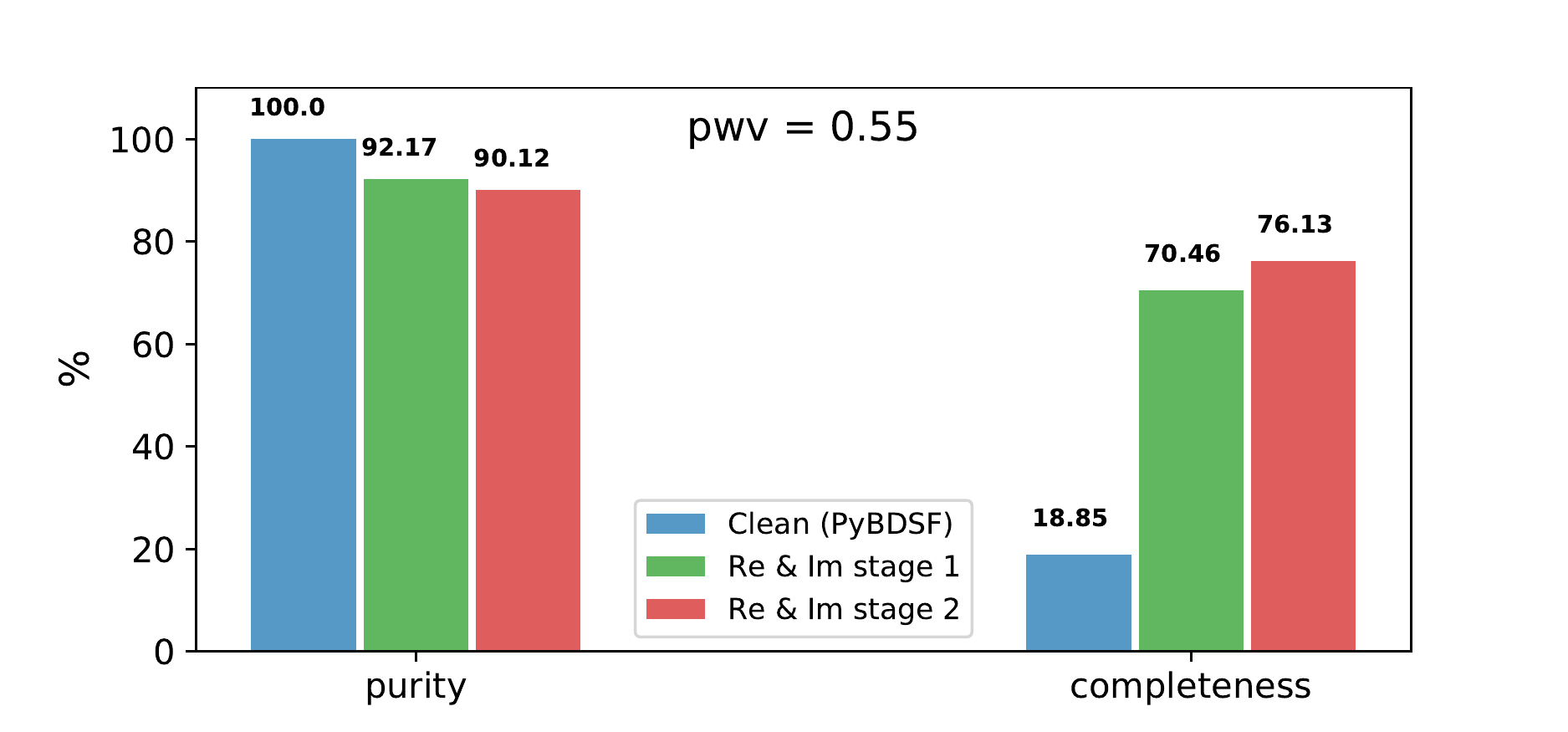}
    \end{subfigure}    
    \hfill
    \begin{subfigure}[c]{.32\linewidth}
      \centering
      \includegraphics[width=1\linewidth]{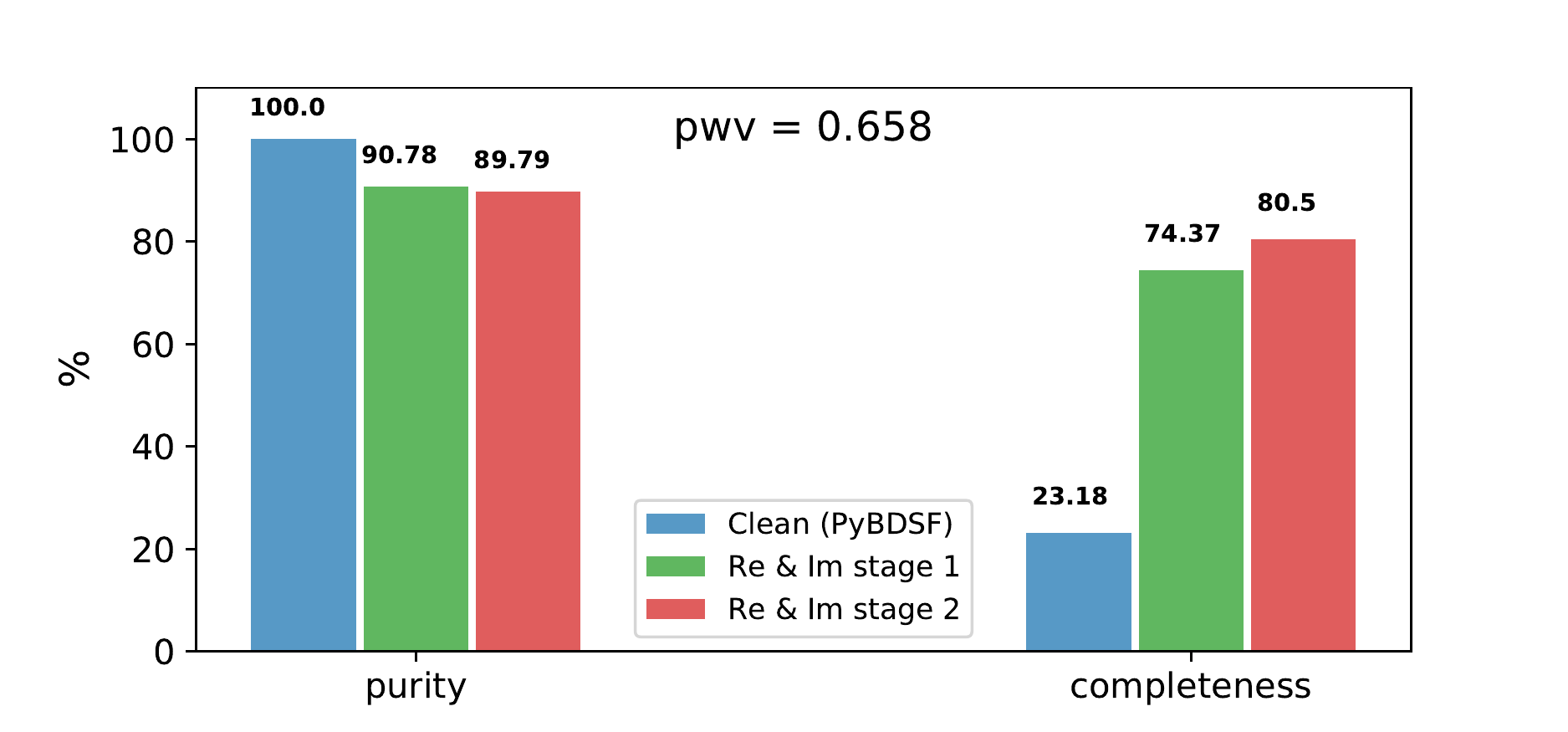}
    \end{subfigure}
    \\
    \begin{subfigure}[c]{.32\linewidth}
      \centering
      \includegraphics[width=1\linewidth]{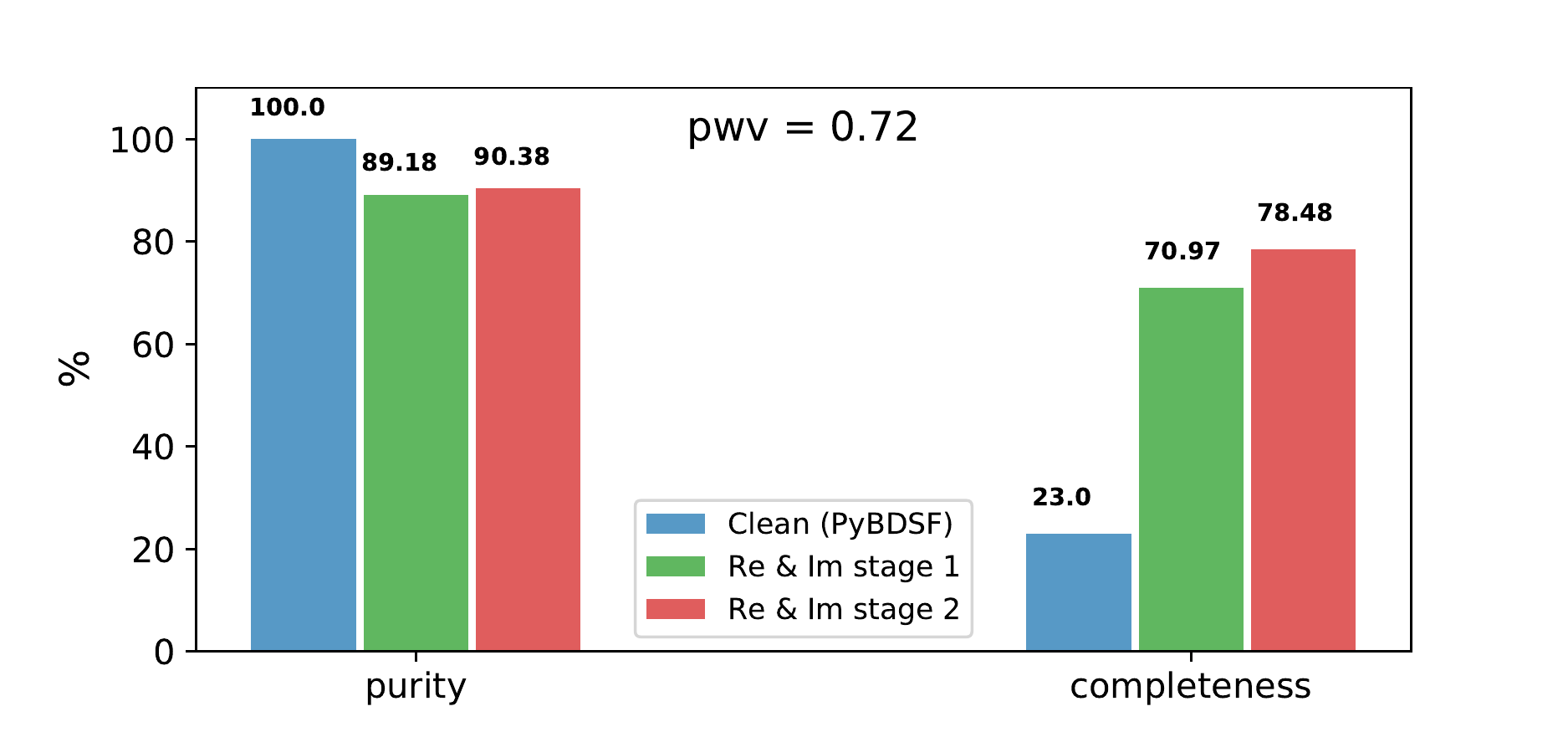}
    \end{subfigure}    
    \hfill
    \begin{subfigure}[c]{.32\linewidth}
      \centering
      \includegraphics[width=1\linewidth]{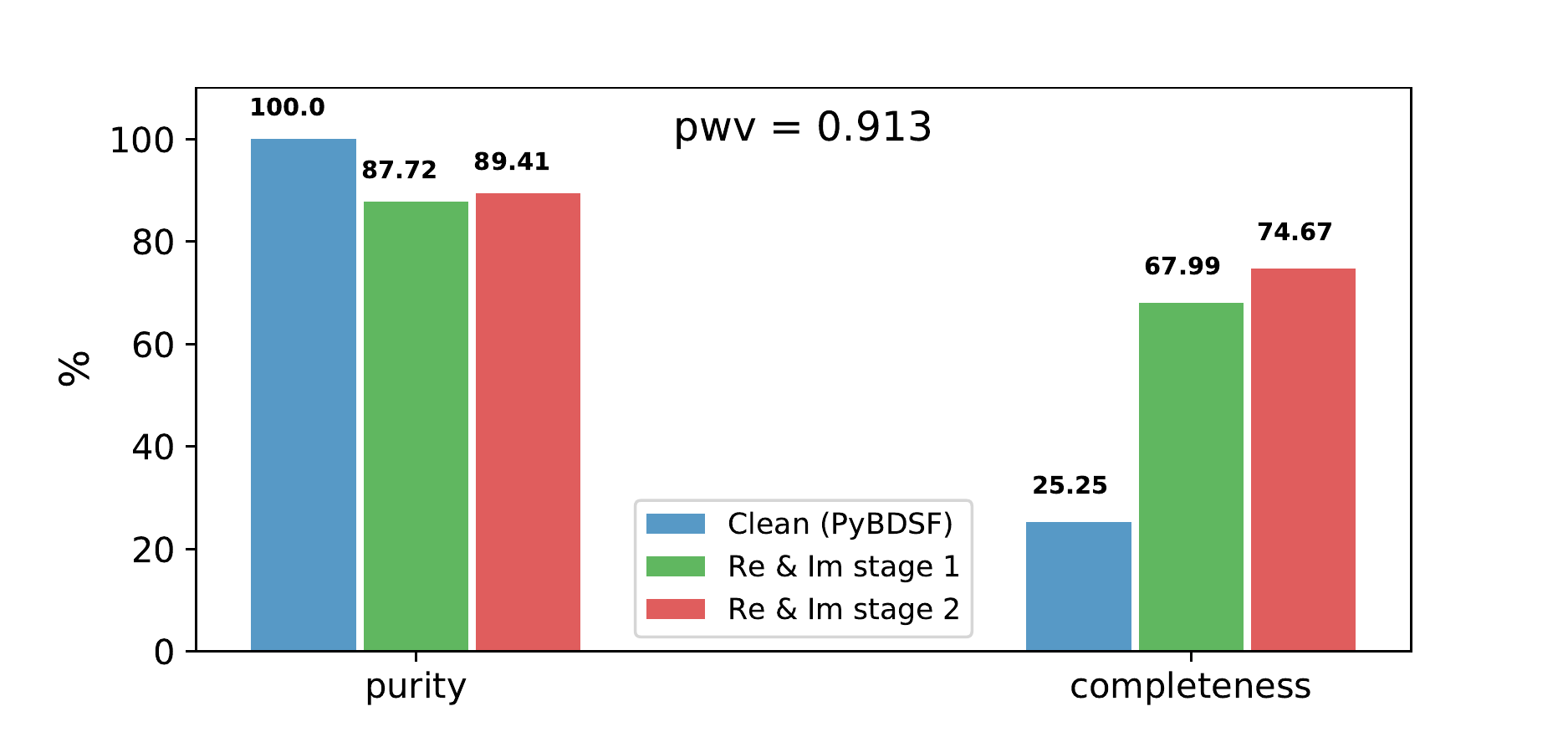}
    \end{subfigure}
    \hfill
    \begin{subfigure}[c]{.32\linewidth}
      \centering
      \includegraphics[width=1\linewidth]{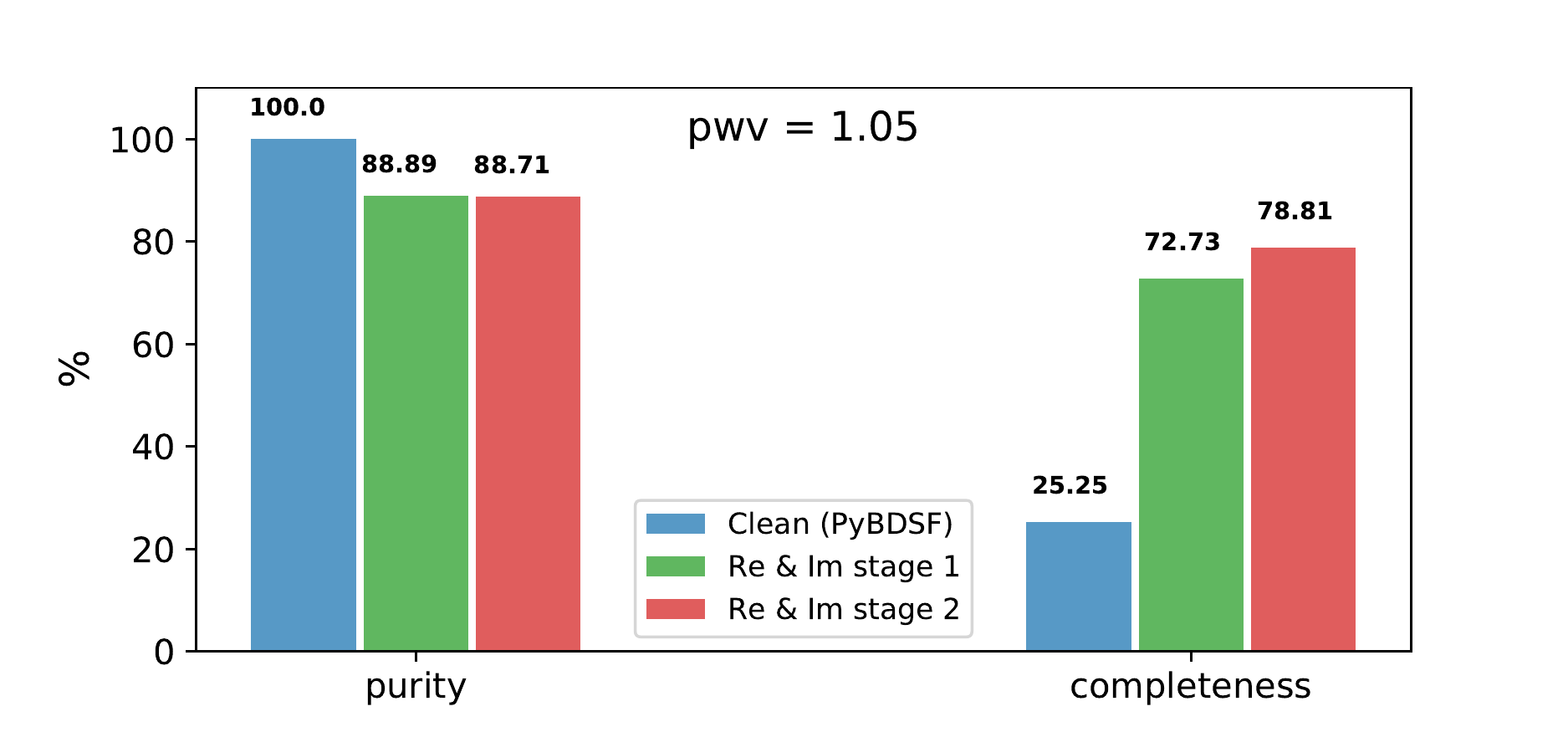}
    \end{subfigure}
    \\
    \begin{subfigure}[c]{.32\linewidth}
      \centering
      \includegraphics[width=1\linewidth]{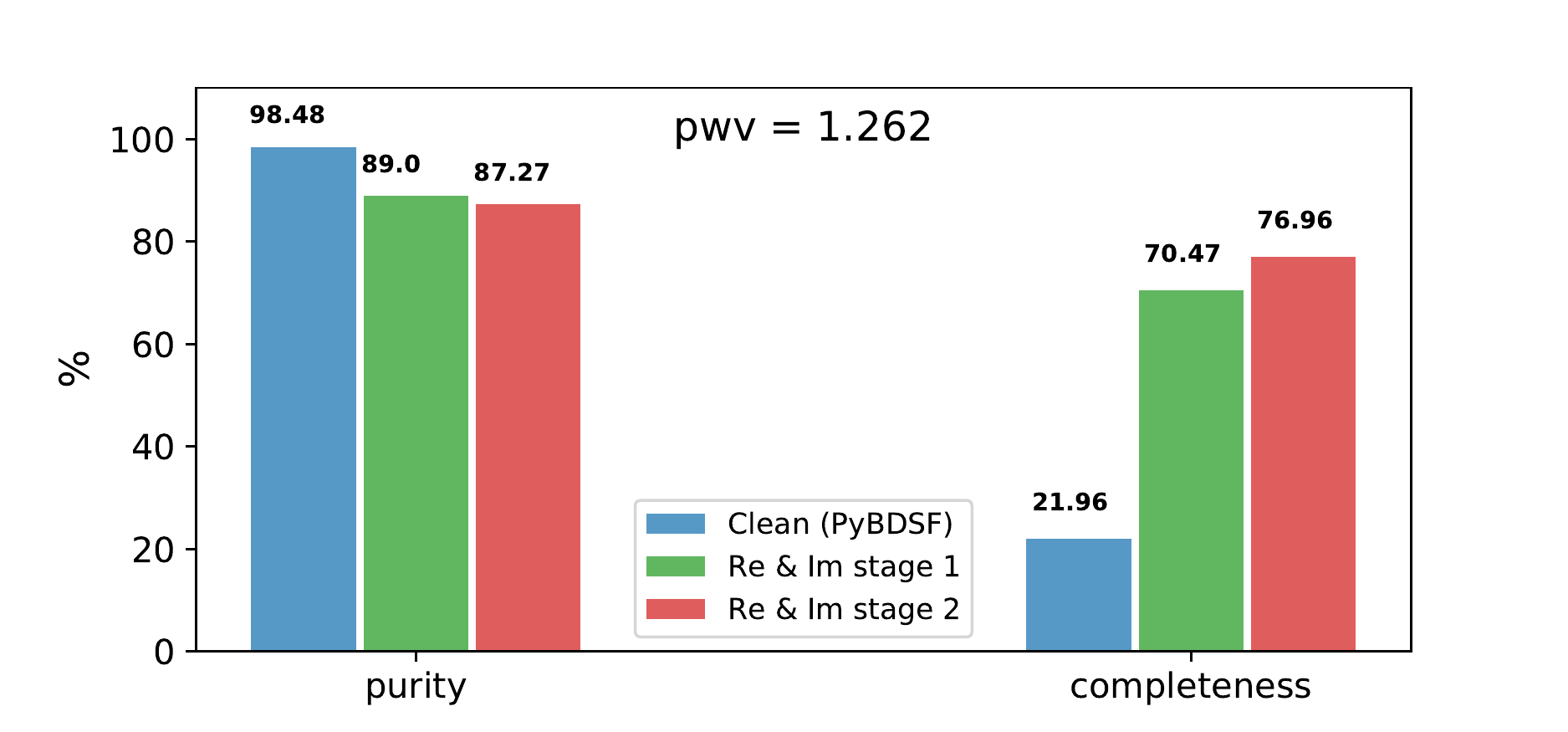}
    \end{subfigure}
    \hfill
    \begin{subfigure}[c]{.32\linewidth}
      \centering
      \includegraphics[width=1\linewidth]{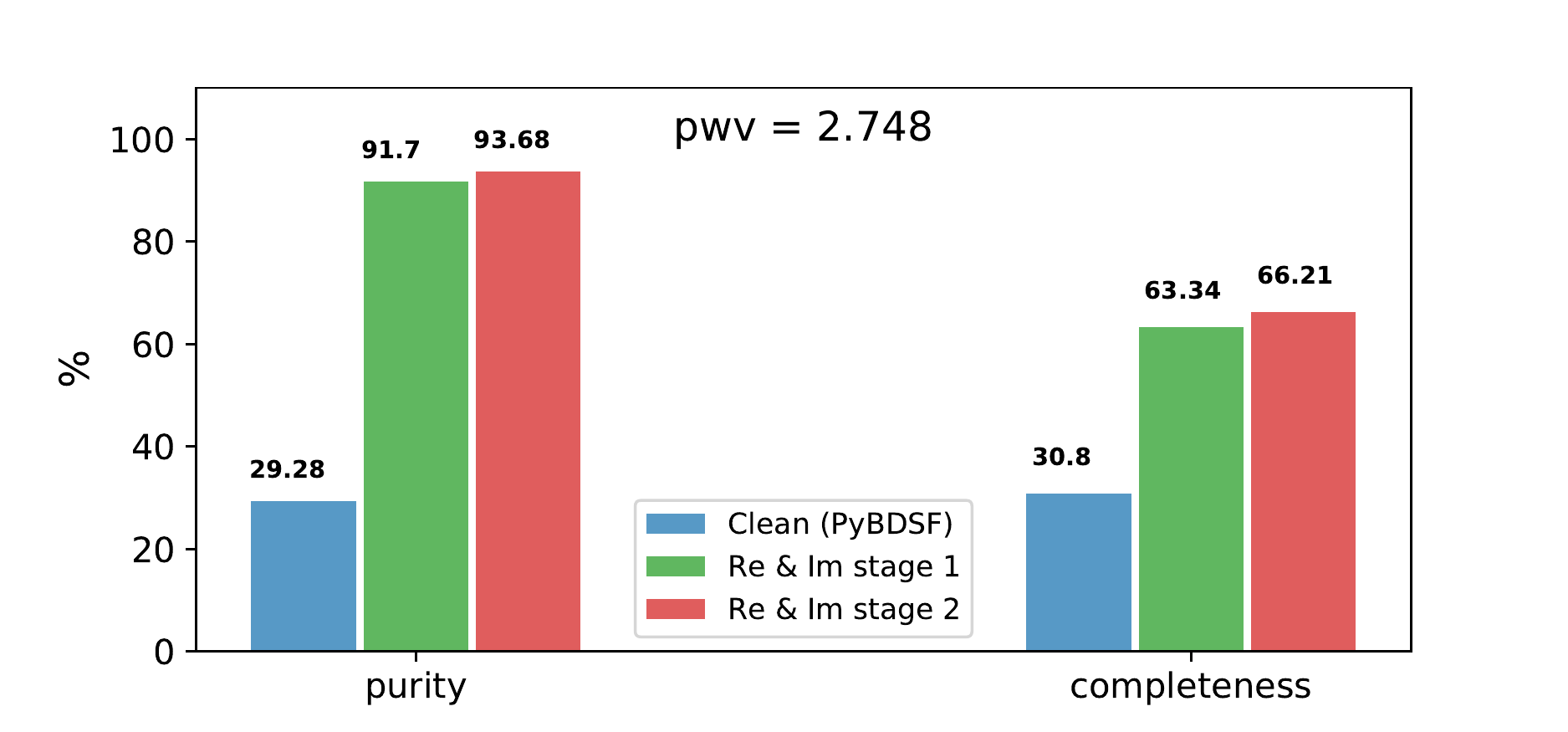}
    \end{subfigure}  
    \hfill
    \begin{subfigure}[c]{.32\linewidth}
      \centering
      \includegraphics[width=1\linewidth]{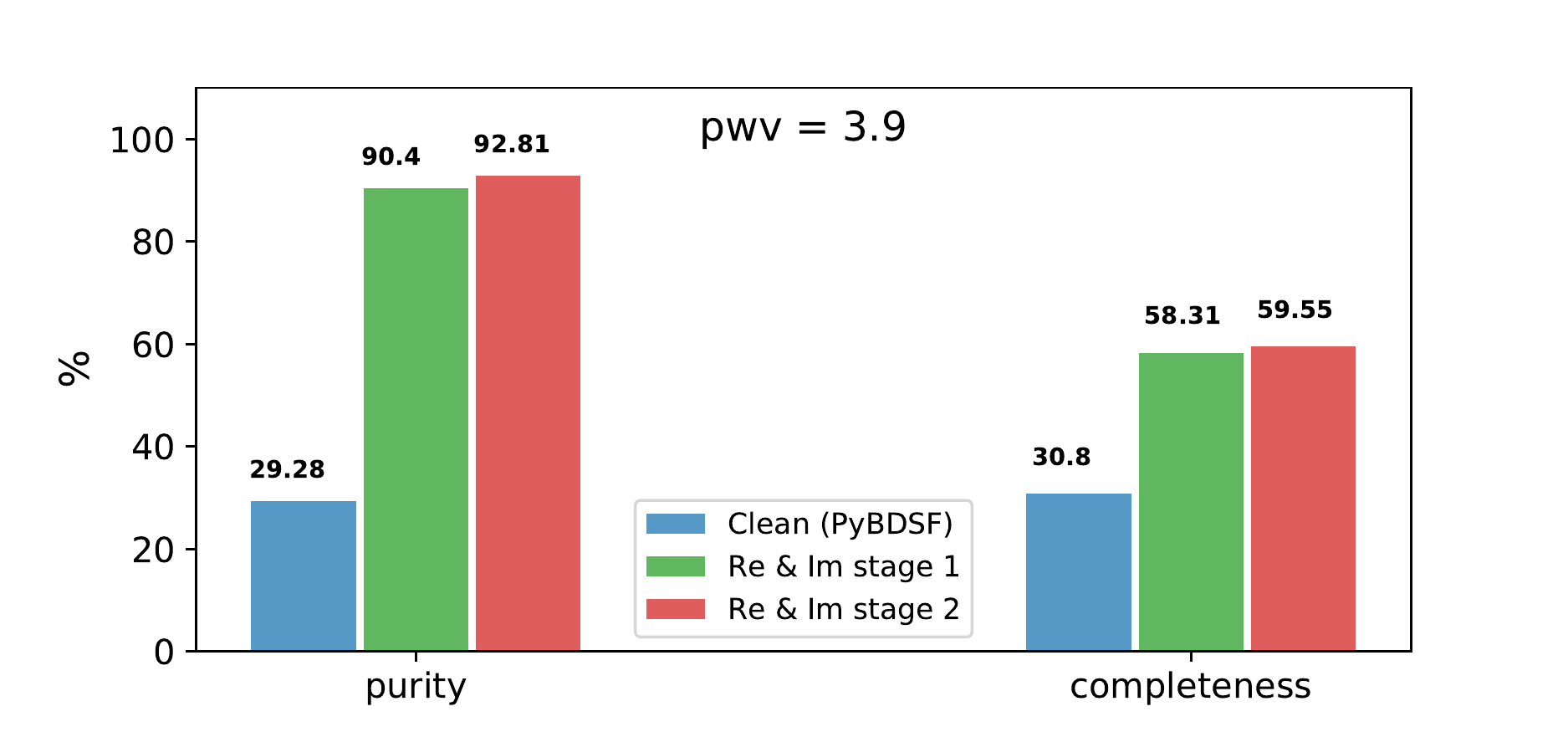}
    \end{subfigure}
    \\
    \begin{subfigure}[c]{.32\linewidth}
      \centering
      \includegraphics[width=1\linewidth]{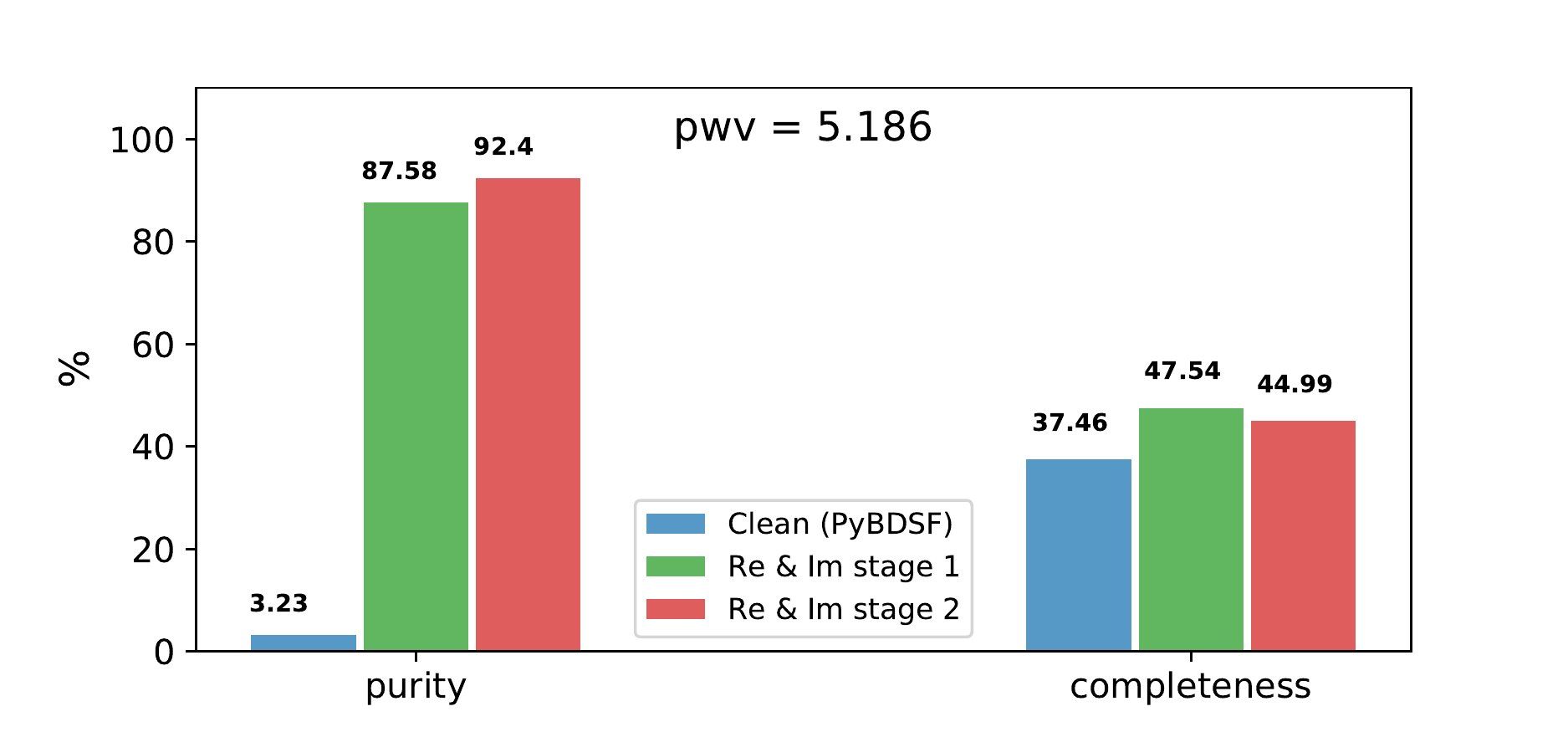}
    \end{subfigure}     
    \hfill
    \begin{subfigure}[c]{.32\linewidth}
      \centering
      \includegraphics[width=1\linewidth]{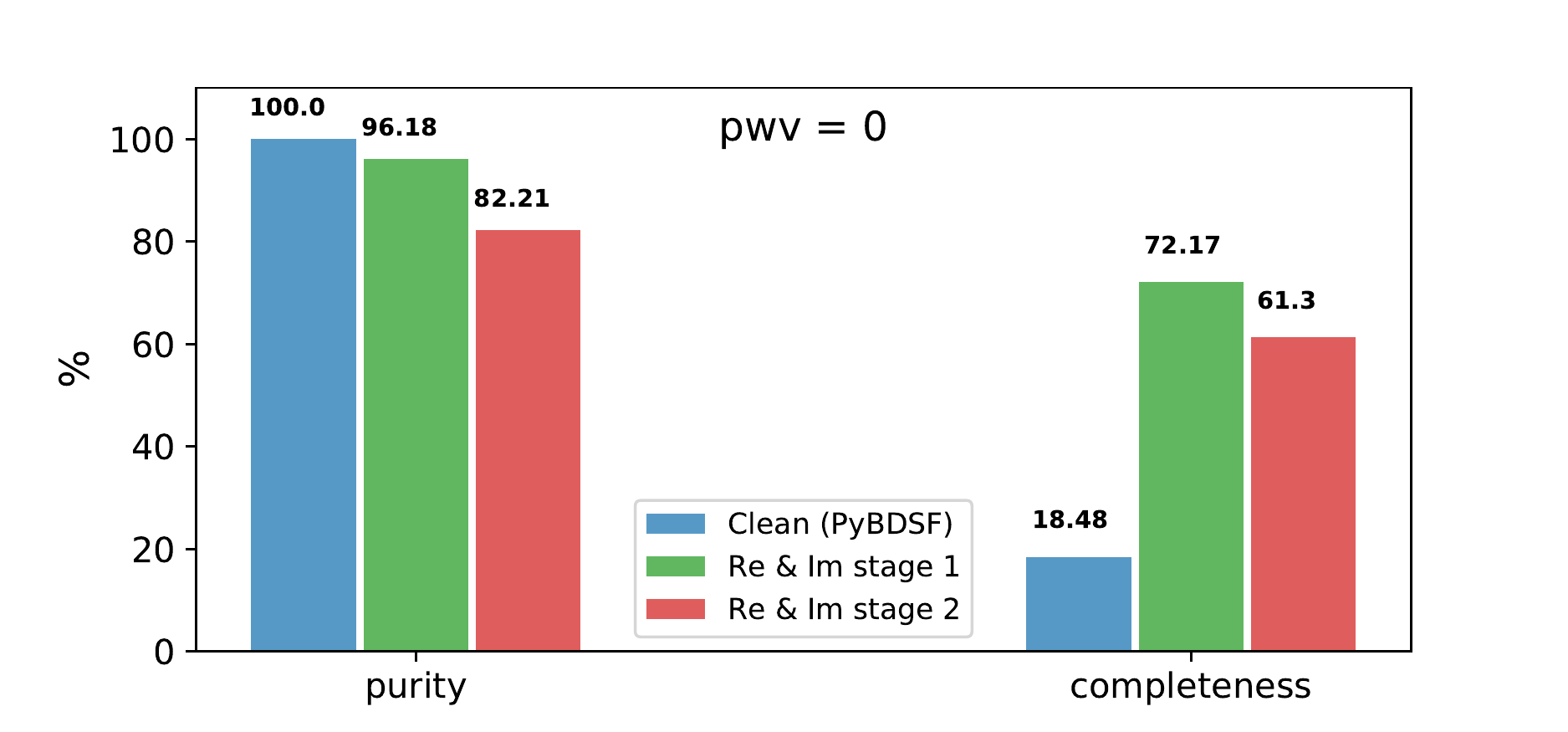}
    \end{subfigure}    
    \caption{The purity and completeness of \pybdsf{} on CLEAN data and the proposed \reim{} framework for the different noise levels, averaged over all S/Ns. From top left to bottom right: \textsc{pwv} = 0.472; \textsc{pwv} = 0.55; \textsc{pwv} = 0.658; \textsc{pwv} = 0.72; \textsc{pwv} = 0.913; \textsc{pwv} = 1.05; \textsc{pwv} = 1.262; \textsc{pwv} = 2.748; \textsc{pwv} = 3.9; \textsc{pwv} = 5.186; \textsc{pwv} = 0.}
    \label{fig:pwv statistic}
\end{figure*}


The results obtained on our simulated data are shown in Fig. \ref{fig:snr2 results}. 
First, we note that the behavior of PyBDSF on our CLEAN and dirty images is similar to what is seen for the \textit{find\_peak} source detector from \textit{astropy} \citep{robitaille2013astropy} used by \citet{bethermin2020alpine}, especially at $f_{\rm norm} \la 4$. Most importantly, for all normalized fluxes  $f_{\rm norm} \la 5$ the proposed framework achieves significantly higher completeness than classical methods. Finally, for high normalized fluxes (and high S/N), the completeness obtained by the proposed framework and \citet{bethermin2020alpine} is similar. From this comparison, we conclude that the proposed method is not only expected to significantly speed up, but also to strongly improve source detection in interferometric imaging.

\subsection{Impact of the noise}

As mentioned in Sect. \ref{sec:data}, in the ALMA setup under investigation, there are two sources of the noise. The first one is the ALMA receiver noise that is fixed in our case, since we fixed the total integration time and observing band. The second one is the atmospheric noise related to the water vapor. To investigate the sensitivity of the proposed framework to the change of noise statistics, we consider the change of the atmospheric noise by varying the \textsc{pwv} parameter. We chose ten different \textsc{pwv} values from 0.472 to 5.186. These values were chosen from the ALMA technical handbook\footnote{\url{https://almascience.nrao.edu/proposing/technical-handbook}} and represent the typical values observed at the ALMA site. For each \textsc{pwv} value, we simulated 100 test samples. Then we tested the proposed \reim{} framework without retraining and without any change in the test parameters. We performed the comparison with \pybdsf{} that is also used without any change of parameters for a fair comparison. The obtained purity and completeness are shown in Fig. \ref{fig:pwv statistic}. It is important to notice that, on average, for the \reim{} framework, we can observe the same dynamic as represented in Sect. \ref{sec:results on noisy data}; namely, stage 2 is slightly superior to stage 1. The performance of \pybdsf{} under the noise \textsc{pwv} smaller than in the main simulations (i.e., \textsc{pwv} = 1.796) is similar to the one reported in Sect.  \ref{sec:results on noisy data}. With the increase in the noise we can observe a drastic decrease in purity from 98 - 100 \% to 30 \% and to 3 \% for the \textsc{pwv} = 5.186. This shows the need for expert knowledge in adapting the parameters. For the proposed \reim{} framework, there is also decrease in the performance but it is not so drastic. On average, the purity remains about 87 - 90 \%, while the completeness decreases from about 60 - 65 \% to 45 -47 \% --  this is 10 \% better than the results from \pybdsf{}. 

In addition, we investigated an extreme case that involves the performance obtained on the noise-free data (\textsc{pwv} = 0) with the models trained on the noisy data. The obtained results are shown in Fig. \ref{fig:pwv statistic} (bottom right). In general, the behavior of the proposed \reim{} framework is quite good. One can observe a certain decrease in the performance at stage 2, while the purity and completeness obtained at stage 1 are high. The completeness obtained by \pybdsf{} is smaller than in case of noisy data. This can be explained by the choice of parameters, assuming the presence of noise.
 
\subsection{Caveats}

It should be pointed out that, despite the advantages discussed above, the proposed framework also has certain limitations. Although it requires less expert knowledge, the amount of training data increases with increasing data complexity, such as the complexity of the shape of the source and their variability, the increase in the number of sources in the field of view, the proximity of the sources with different S/N.

Moreover, taking into account that the proposed framework shows  high sensitivity to the low S/N sources, it might lead to false sources in the source free sky models (e.g., pure background noise). And although such a scenario usually is not considered among state-of-the-art approaches, we tested the proposed \reim{} framework on 1000 source free noisy sky models. After stage 1, the false sources are detected in about 40 \% of the sky models, while after stage 2, only in 27 \% of cases. The results obtained can be explained by the fact that the source free sky models were not taken into account during the framework training.

\subsection{Future developments and applications}

The proposed framework can be applied to real data, and will be tested with available data from the ALMA archive,  such as data from \acosmos \citep{liu19a3cosmos}. This will, in particular, also allow us to examine the behavior with data taken in different conditions and accounting for all noise sources present in the system.

The next tasks to be tackled from uv-data alone will include source characterization that are measurements of fluxes, source morphologies, and others, as well as the treatment of extended sources, possibly with complex morphologies.
New machine learning-based approaches on visibilities should also be able to handle spectral lines and thus be applicable to an even broader range of astrophysical questions.
In parallel, our framework should also be tested and generalized to cover a wide range of interferometric data taken with very different facilities and across a wide spectral range from the millimeter to the radio domain. If proven successful, machine learning based methods could have a strong impact on our future handling of interferometric data and help push the discovery space of upcoming observatories even further.

\section{Conclusions}
\label{sec:conclusions}

Radio astronomy is at a historic moment in its development. Innovative DNN-based methods may drastically change the way measurements are processed and, thus,  may also improve our capacity to detect faint and complex signals. In this context, we have taken a new ML-based approach to solve the problem of source localization, directly working on the natural measurement sets (visibilities) and without reconstructing the dirty or CLEAN image.
The proposed framework consists of two stages generating source localization maps (see Fig. \ref{fig:main model}). 

To train the network and then validate and test the proposed framework, we used synthetic data generated with the CASA data processing software \citep{mcmullin2007casa}, which is the software tool for interferometric data/observations with ALMA and similar observatories, and \pybdsf{} \citep{mohan2015pybdsf} for the classical source detection.
The sky simulations generated in this work were chosen to represent typical extra-galactic (sub)-millimeter continuum survey observations 
undertaken with ALMA. The simulations included both ideal noise-free and realistic noisy simulations.
The comparisons between the proposed DNN-based approach and existing state-of-the-art source localization pipelines can be summarized as follows:
\begin{itemize}
    \item In ideal noise-free cases, the proposed framework has overall the same purity and completeness as the traditional pipeline using PyBDSF source localization.
    \item For noisy data, the proposed DNN-based method fares significantly better than classical source localization algorithms.
    For example, the \reim{} framework shows  completeness that is more than three times as high as it is for the same purity as the traditional PyBDSF-based pipeline 
    for all sky models, when considering a uniform, random distribution of sources with S/N$=1-10$ (Fig.~\ref{fig:snr sources statistic}).
    \item In the low-S/N regime (S/N $\la 5$), the performance gain of the proposed method is very high: while the traditional pipeline achieves a completeness of 2-5 \% for S/N $<3$, the proposed framework detects sources with a completeness of $\sim 20$ ($45-55$) \%  for S/N = 2 (3).
    \item The new source-detection method represents an important gain in execution time, with total execution times that are more than 30 times faster than the traditional pipeline that involves image reconstruction from the uv-plane and source detection (Table \ref{tab:exec time}).
\end{itemize}

We have investigated the impact of different factors on the efficiency of the proposed framework, in particular:
\begin{itemize}
    \item {\em Input data representation:}
    While traditional DNNs are designed to work with real-valued data, we have tested representations of the complex-valued uv-data by real-valued real and imaginary or magnitude and phase components. The obtained results show that the real and imaginary components are better better suited to the proposed framework, since they have the same dynamic range in contrast to the magnitude-phase case.
    \item %
    {\em Source density:}
    Traditional source detection in the image plane demonstrates a stable behavior that does not depend much on the number of sources. However, in the proposed framework the completeness decreases with increasing numbers of sources (Fig. \ref{fig:snr performance by sources}). Despite this, in tested conditions which represent typical ALMA extra-galactic continuum surveys, the proposed framework reaches a higher completeness than traditional source detection algorithms.
\end{itemize}

In short, we have developed a DNN-based method of source detection using only uv-plane observations. We have shown that it provides strong improvements in detecting sources in the low-to-intermediate-S/N regime.
The new approach can already be applied to existing interferometric observations and it opens many new possibilities which will be explored in the near future. Machine learning-based methods have the potential to significantly alter our approach to interferometric data, as we enter the era of new facilities like the Square Kilometer Array\footnote{\url{https://www.skao.int/}} (SKA).


\begin{acknowledgements}

We acknowledge the referee for their comments. O. Taran and O. Bait are supported by the {\em AstroSignals} Sinergia Project funded by the Swiss National Science Foundation. 

We also thank B. Magnelli and M. Béthermin, and the nordic arc node \url{https://nordic-alma.se}.

\end{acknowledgements}

%
%

\bibliographystyle{aa} 
\bibliography{references}

\begin{appendix} 

\section{UV Coverage and the dirty beam}
\label{appendix: uvcoverage}

\begin{figure}[ht!]
    \centering
    \begin{subfigure}[c]{0.9\linewidth}
      \centering
      \hspace{-1.25cm}
      \includegraphics[width=0.7\linewidth]{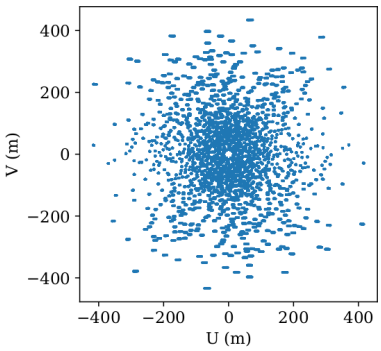}
    \end{subfigure}
    \\
    %
    \begin{subfigure}[c]{0.9\linewidth}
      \centering
      \includegraphics[width=0.65\linewidth]{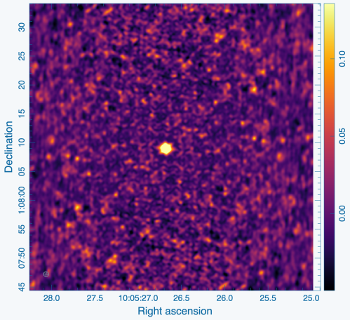}
    \end{subfigure}
    \caption{Example of the total UV coverage for one of our ALMA 12-m array simulation pointing lasting 20 mins with 240 channels (top). The corresponding dirty beam is shown in the bottom panel.}
    \label{appendix fig: uvcoverage dirty beam}
\end{figure}

Figure \ref{appendix fig: uvcoverage dirty beam} shows the typical UV coverage for our simulations and the corresponding dirty beam. Note: the dirty beam has several sidelobes, but they are of a smaller amplitude, owing to the relatively filled UV. 

\section{LESTA computing cluster and simulation details}
\label{apppendix: lesta}

The simulations presented in this work are produced on the LESTA computing cluster hosted at the University of Geneva Department of Astronomy. The details of the used resources:
\begin{itemize}
    \item 16 Intel Xeon (E5-2640 v3, 2.6GHz) with 128GB (2133 MHz) memory (per node)
    \begin{itemize}
        \item number of nodes: 32;
        \item number of cores: 512
    \end{itemize}
    \item 32 Intel(R) Xeon(R) Gold 5218 CPU @ 2.30GHz with 256GB (2666 MHz) memory (per node):
    \begin{itemize}
        \item number of nodes: 8;
        \item number of cores: 256
    \end{itemize}
\end{itemize}

 Each simulation or pointing was run in parallel using slurm job arrays on both resources, depending on availability. Each simulation takes approximately 12 minutes.

\section{Sub-sampling}
\label{apppendix: dnn sampling}

It is important to note that networks are quite demanding for input data stability such as the data size, the data dynamic range, and so on. In the considered antenna configuration, there are 50 antennas. That gives 1225 measurements. Taking into account the Earth's movement during the integration time, along with the source positions and the subsequent gridding, the final number of measured frequencies (i.e., dimensionality of the input data) increases and is variable from about 1300 until about 1500. To solve the issue with the variable size of the input data, we chose 1400 most frequently measured uv-positions. To do this,  we  first took 100 gridded data at random and converted them to the binary maps, where unmeasured frequencies were set to 0 and the measured ones to 1, disregarding the real intensity. Then we summed up these 100 binary maps and chose 1400 positions with the highest values sorted in the decreasing order. That, in its turn, leads to the final binary map with 1400 values equal to 1 and the others equal to 0. Then every gridded piece of data was multiplied by this final binary map. It is clear that in the cases when the gridded data contain more than 1400 measured frequencies, it leads to the lost of information. However, we allowed for this in order to have the fixed size input data for the network.

\section{Proposed framework training }
\label{apppendix: training of proposed framework}

The Python implementation of the proposed framework is be publicly available at \url{https://github.com/taranO/ml-based_source_localization_from_uv-plane}.

For the training of the proposed framework, the data set was split into the train ($80 \%$), validation ($10 \%$), and test subsets ($10 \%$). The training was done three times under three different seeds. \\

\noindent \textbf{Stage 1}
\begin{table}[th!]
   \renewcommand{\arraystretch}{1.5}%
    \centering
    \caption{Architecture of the $g_\btheta$ model for $\bar{\y} \in \mathbb{R}^{2 \times K}$ represented by the real and imaginary components.}
    \label{app tab:stage 1 re im}    
    \begin{tabular}{ccc}
        Size & Layer & Param \# \\ \hline \hline  
         $2 \times 1400$ & Input & \\
         $2 \times 262144$ & Dense, ReLU & 367263744 \\
         $512 \times 512$ & Reshape & \\
         $512 \times 512$ & Weighted sum, Tanh & 524288 \\ \hline
    \end{tabular}
\end{table}
\vspace{-0.5cm}
\begin{table}[th!]
   \renewcommand{\arraystretch}{1.5}%
    \centering
    \caption{Architecture of the $g_\btheta$ model for $\bar{\y} \in \mathbb{R}^{2 \times K}$ represented by the magnitude and phase components.}
    \label{app tab:stage 1 mag ph}    
    \begin{tabular}{ccc}
        Size & Layer & Param \# \\ \hline \hline  
         $2 \times 1400$ & Input & \\
         $512 \times 512$ & Weighted mult. & 2800 \\
         $2 \times 262144$ & Dense, ReLU & 367263744 \\
         $512 \times 512$ & Reshape & \\
          \hline
    \end{tabular}
\end{table}

\textit{Training} of the proposed framework was performed on the NVIDIA Titan RTX GPU with 24GB memory during 1000 epochs (the training time is about 3 hours) with the learning rate 1e-4 and the batch size of 312. 

\begin{table}[th!]
   \renewcommand{\arraystretch}{1.5}%
    \centering
    \caption{Architecture of the $q_\bphi$ model.}
    \label{app tab:stage 2}    
    \begin{tabular}{ccc}
        Size & Layer & Param \# \\ \hline \hline  
        $512 \times 512 \times 1$ & Input & \\
        $256 \times 256 \times 8$ & Conv2D, ReLU & 80 \\
        $256 \times 256 \times 8$ & BN & 32 \\
        $128 \times 128 \times 16$ & Conv2D, ReLU & 1168 \\
        $256 \times 256 \times 16$ & BN & 64 \\
        $256 \times 256 \times 8$ & Conv2DTran, ReLU & 1160 \\
        $256 \times 256 \times 8$ & BN & 32 \\  
        $512 \times 512 \times 1$ & Conv2DTran, Sigmoid & 73 \\
        $512 \times 512 \times 1$ & BN & 4 \\ \hline
    \end{tabular}
    
\end{table}

\noindent\textit{Stage 2 training} was performed on the NVIDIA GeForce RTX 2080 Ti GPU with 11GB memory during 500 epochs (the training time is about 8 hours)  with the learning rate 1e-3 and the batch size of 32.

\begin{figure}[ht!]
      \centering
      \includegraphics[width=0.95\linewidth]{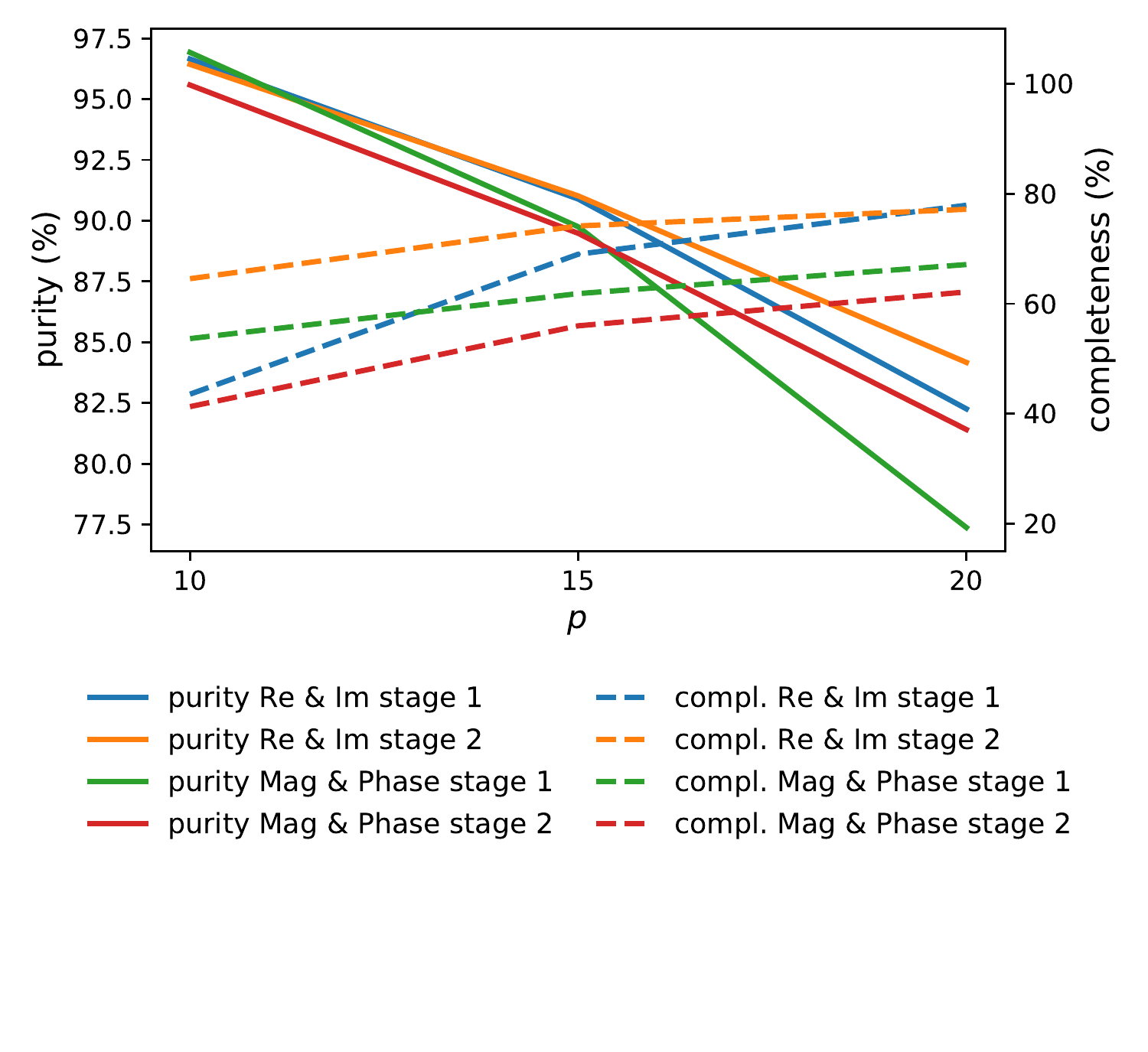}
      \vspace{-1.5cm}
      \caption{Dependence of the purity and completeness on the limit on the number of detected sources $p$.}
      \label{appendix fig: jump_lim}
\end{figure}

\textit{Post-binarization}: we do not apply the fixed threshold $t$ to binarize the DNN outputs $\hat{\m}^{\in{N \times M}}$. Instead, the binarization threshold $t$ is chosen on the fly for each particular $\hat{\m}$. From the prior knowledge, we know that there are no more than five sources in each simulation. To determine the optimal threshold, $t$, the proposed method performs an iterative thresholding until the number of detected sources, $p$, both true and false, is not bigger than 15. This value was chosen empirically based on the trade-off between the corresponding purity and completeness as shown in Fig. \ref{appendix fig: jump_lim}.

\section{State-of-the-art source detectors}
\label{appendix: sota detectors}

\subsection{Noise-free data} 
\label{appendix: sota detectors noise-free} 

\begin{table}[ht!]
    \small
    \renewcommand{\arraystretch}{1.5}%
    \centering
    \caption{Performance (in \%) of the state-of-the-art source detectors on the CLEAN images in the noise-free case.}    
    \begin{tabular}{lccc} \hline
         & PyBDSF & AEGEAN & PySE \\ \hline 
         Purity       & 94.21 ($\pm0.55$) & 94.41 ($\pm0.45$) & 94.23 ($\pm0.8$)\\
         Completeness & 93.07 ($\pm0.11$) & 93.62 ($\pm0.13$) & 91.49 ($\pm0.25$) \\ \hline
    \end{tabular}
    \label{tab:noise-free sota comparison}
\end{table}

A comparison of the performance of the state-of-the-art PyBDSF, AEGEAN \citep{Hancock12, Hancock18}, and PySE \citep{carbone2018pyse} methods on our data set is given in Table \ref{tab:noise-free sota comparison}. The parameters of the detectors were chosen to provide the purity of about 94 \%. It is easy to see that the purity and completeness obtained for \pybdsf{} and \aegean{} are very similar. Under the same purity, \pyse{} has slightly smaller completeness. 

\subsection{Noisy data} 
\label{appendix: sota detectors noisy}

\begin{table}[ht!]
    \small
    \renewcommand{\arraystretch}{1.5}%
    \centering
    \caption{Performance (in \%) of the state-of-the-art source detectors on the CLEAN images in the noisy case, averaged over all S/Ns.}    
    \begin{tabular}{lccc} \hline
                & PyBDSF & AEGEAN & PySE \\ \hline
         Purity & 90.60 ($\pm1.06$) & 48.49 ($\pm1.16$) & 65.73 ($\pm1.56$)\\
         Completeness & 25.48 ($\pm0.42$) & 25.57 ($\pm0.60$) & 25.64 ($\pm0.29$)\\ \hline
    \end{tabular}

    \label{tab:noisy sota comparison}
\end{table}

Similarly to the noise-free case, we compare the performance of the \sota{} detectors PyBDSF, AEGEAN, and PySE. The obtained purity and completeness are given in Table \ref{tab:noisy sota comparison}. The parameters of the detectors were chosen to provide the maximum purity under the assumption that the  completeness should be no less than 25 \%. In contrast to the noise-free case, in case of noisy data under the same completeness \pyse{} outperforms \aegean{} in terms of the purity. \aegean{} detector has almost two times smaller purity compared to \pybdsf{}. \pybdsf{} demonstrates the best performance. 

\end{appendix}

\end{document}